\documentclass[fleqn,usenatbib]{mnras}

\usepackage{newtxtext,newtxmath}

\usepackage[T1]{fontenc}

\DeclareRobustCommand{\VAN}[3]{#2}
\let\VANthebibliography\thebibliography
\def\thebibliography{\DeclareRobustCommand{\VAN}[3]{##3}\VANthebibliography}

\usepackage{graphicx}	
\usepackage{amsmath}	

\def\hii{H\,\textsc{ii}}

\def\oii{O\,\textsc{ii}}
\def\oiii{O\,\textsc{iii}}
\def\sii{S\,\textsc{ii}}
\def\siii{S\,\textsc{iii}}

\def\ciii{C\,\textsc{iii}}

\def\heii{He\,\textsc{ii}}

\def\neiii{Ne\,\textsc{iii}}

\def\ariii{Ar\,\textsc{iii}}

\def\siiii{Si\,\textsc{iii}}
\def\feiii{Fe\,\textsc{iii}}

\def\buv{$\beta_{\mathrm{UV}}$}
\def\met{$12+\log(\mathrm{O/H})$}

\usepackage{lineno}

\title[$\alpha$-element, C, and Fe abundances]{
JADES: the chemical enrichment pattern of distant galaxies -- \\$\alpha$ enhancement, silicon depletion, and iron enhancement
}

\author[Y. Isobe et al.]{Yuki Isobe,$^{1,2,3}$\thanks{E-mail: yi264@cam.ac.uk (YI)}
Roberto Maiolino,$^{1,2,4}$
Xihan Ji,$^{1,2}$
Francesco D'Eugenio,$^{1,2}$
Charlotte Simmonds,$^{1,2}$
\newauthor
Jan Scholtz,$^{1,2}$
Ignas Juod\v{z}balis,$^{1,2}$
Aayush Saxena,$^{5,4}$
Joris Witstok,$^{6,7}$
Chiaki Kobayashi,$^{8}$
Irene Vanni,$^{9}$
\newauthor
Stefania Salvadori,$^{9,10}$
Kuria Watanabe,$^{11,12}$
Stephanie Monty,$^{13,14,15}$
Vasily Belokurov,$^{15}$
Anna Feltre,$^{10}$
\newauthor
William McClymont,$^{1,2}$
Sandro Tacchella,$^{1,2}$
Mirko Curti,$^{16}$
Hannah \"Ubler,$^{17}$
St\'ephane Charlot,$^{18}$
\newauthor
Andrew J.\ Bunker,$^{5}$
Jacopo Chevallard,$^{5}$
Emma Curtis-Lake,$^{8}$
Nimisha Kumari,$^{19}$
Pierluigi Rinaldi,$^{20}$
\newauthor
Brant Robertson,$^{21}$
Christina C.\ Williams,$^{22}$
and
Chris Willott$^{23}$
\\
\\
Affiliations are listed at the end of the paper.
}

\date{Accepted 2026 January 12. Received 2025 December 15; in original form 2025 September 22}

\pubyear{\the\year{2026}}

\begin{document}
\label{firstpage}
\pagerange{\pageref{firstpage}--\pageref{lastpage}}
\maketitle

\begin{abstract}
We present gas-phase abundances of carbon (C), $\alpha$-elements (O, Ne, Si, and Ar) and iron (Fe) obtained from stacked spectra of high-$z$ star-forming galaxies
with the deep Near Infrared Spectrograph medium-resolution data from the James Webb Space Telescope Advanced Deep Extragalactic Survey.
Our 564 sources at $z=4$--7 have a median stellar mass of $\log(M_{*}/M_{\odot})=8.46$ and a median star-formation rate of $\log(\mathrm{SFR}/M_{\odot}\,\mathrm{yr^{-1}})=0.30$, placing them close to the star-formation main sequence.
We find that the stacked spectrum of all our 564 sources has relatively low
$[\mathrm{C/O}]=-0.70$,
moderate
$[\mathrm{Ne/O}]=-0.09$,
and low
$[\mathrm{Ar/O}]=-0.28$
values at a low gas-phase metallicity of $12+\log(\mathrm{O/H})=7.71$ ($Z\sim 0.1~Z_\odot$), suggesting dominant yields of core-collapse supernovae evolved from massive stars.
The detection of a weak \siiii] emission line in our stacked spectrum provides a silicon-to-oxygen abundance ratio of
$[\mathrm{Si/O}]=-0.63$,
which is lower than that of
stars in the Milky Way disc and lower than expected by chemical evolution models,
suggesting silicon depletion onto dust grains.
Likewise, this Si/O value is lower than that we newly derive for two individual $z>6$ galaxies (GN-z11 and RXCJ2248) with negligible dust attenuation.
By performing spectral stacking in
bins of $M_{*}$, SFR, specific SFR (sSFR), and ultra-violet (UV) continuum slope \buv,
we identify [\feiii] line detections in the high-sSFR bin and the blue-\buv\ bin, both of which exhibit supersolar Fe/O ratios, while their C/O, Ar/O, and Si/O ratios are comparable to those of the all-sources stack.
Our findings support a chemically young gas composition with rapid dust depletion in the general population of high-$z$ star-forming galaxies, while raising the possibility of anomalous, selective Fe/O enhancement at the very early epoch of star formation.
\end{abstract}




\section{Introduction} \label{sec:intro}
Chemical abundances of galaxies provide key insights into star formation and galaxy evolution.
The first generation of stars formed from hydrogen (H) and helium (He) synthesised during the Big Bang \citep[e.g.,][]{Cyburt2016}, and subsequently stars of various masses assembled to form galaxies.
These stars contribute to enrich elements more massive than He (so-called metals), with the stellar mass governing which metals are synthesised and ultimately released into the interstellar medium (ISM).
Massive stars with $\sim 8$--100 solar masses ($M_{\odot}$) evolve into core-collapse supernovae (CCSNe; e.g., \citealt{Nomoto2013}) at the end of stellar lifetimes (typically $\sim3$--40 Myr, which slightly depends on metallicity; \citealt{Portinari1998}), ejecting a large amount of carbon (C) and $\alpha$-elements such as oxygen (O) and neon (Ne).
Conversely, after $\sim40$ Myr, dying low-mass stars with $\sim1$--8 $M_{\odot}$ eject their outer layers via stellar winds during the asymptotic giant branch (AGB) phase \citep[e.g.,][]{Herwig2005}.
AGB stars with $\sim4$--7 $M_{\odot}$ contribute to nitrogen (N) enrichment via the CNO cycle, while low-mass AGB stars with $\sim1$--4 $M_{\odot}$ release C \citep[e.g.,][]{Kobayashi2011,Karakas2014}.
White dwarf (WD) remnants emerge after the outer layers are ejected, and sufficient mass accretion from companion stars onto the WDs lead to Type-Ia SNe \citep[e.g.,][]{Maoz2014}, which produce iron (Fe) and heavy $\alpha$-elements such as silicon (Si) and argon (Ar) \citep[e.g.,][]{kob20ia}.
Due to the longer delay time of AGB stars and Type-Ia SNe, C/O, N/O, Si/O, Ar/O, and Fe/O ratios are expected to increase with increasing galaxy age and metallicity \citep[see also][]{Maiolino2019}, as predicted by many chemical evolution models \citep[e.g.,][]{Vincenzo2016,Suzuki2018,Kobayashi2020}.
In fact, many observations for absorption lines of individual stars in the Milky Way (MW) have shown that C/O, N/O, and Fe/O ratios generally increase with metallicity \citep[e.g.,][]{Nicholls2017}, while the observations of individual stars are only limited to the Local Group \citep[e.g.,][]{McWilliam2013}.

Abundance ratios based on emission lines from ionised gas within 
galaxies at low redshifts ($z\sim0$) have been actively investigated \citep[e.g.,][]{Lequeux1979,Izotov1999,Izotov2006}.
Such gas-phase abundances can be affected by dust depletion, in particular, a significantly large fraction of Fe is depleted onto dust grains compared to C, N, and O \citep[e.g.,][]{Whittet2003,Jenkins2009,Roman-Duval2022}.
This explains the general observational trends that show that although the gas-phase C/O and N/O ratios increase with metallicity in the high metallicity regime \citep[e.g.,][]{Pilyugin2012,Berg2019}, the gas-phase Fe/O ratio actually decreases with metallicity \citep[e.g.,][]{Izotov2006,Mendez-Delgado2024}.
Whereas there are a number of interesting exceptions that suggest other contributions \citep[e.g.,][]{Telles2014,Kumari2018}, a general understanding of gas-phase abundances at $z\sim0$ would be as follows: $\alpha$-elements from CCSNe dominate at the early epoch, followed by C and N from AGB stars together with Fe and heavy $\alpha$-elements from Type-Ia SNe \citep[e.g.,][]{Kobayashi2020}, while Fe is depleted onto dust grains.
Indeed, $z\sim0$ dwarf galaxies show $\alpha$-element ratios of Ne/O and Ar/O relatively constant within a wide range of gas-phase metallicity \citep[i.e., $12+\log(\mathrm{O/H})\sim7.2$--8.5;][]{Izotov2006}, while delayed enrichment of Ar, a heavy $\alpha$-element compared to O, has been observed in planetary nebulae \citep{Pottasch2006,Pagomenos2018,Arnaboldi2022}.

It is worth mentioning that C/O and N/O studies based on emission line observations were extended to $z\sim2$--3 \citep[e.g.,][]{Steidel2016,Kojima2017,Hayden-Pawson2022,Llerena2023} prior to the James Webb Space Telescope (\textit{JWST}; \citealt{Gardner2023,Rigby2023}).
However, the launch of this telescope has enabled studies of chemical abundances even at $z>4$ (see Appendix \ref{apsec:ref}).
Using spectroscopic data of the \textit{JWST}/Near Infrared Spectrograph (NIRSpec; \citealt{Ferruit2022,Jakobsen2022}), various studies have shown that several $z>4$ galaxies have C/O and Ne/O ratios comparable to those at $z\sim0$ \citep[e.g.,][]{Arellano2022,Isobe2023c,Tang2025}.
Some studies have reported that some $z\gtrsim4$ galaxies have low C/O ratios \citep[e.g.,][]{Jones2023,Stiavelli2023} and low Ar/O ratios \citep{Bhattacharya2025,
Stanton2025} close to those of CCSNe.
These results are interpreted to imply a gas composition dominated by CCSN yields with little contribution from low-mass stars.
Additionally, it is noteworthy that stellar absorption lines of massive quiescent galaxies at $z=1$--3 show deficits in C and Fe relative to $\alpha$-elements such as Mg, indicating that their gas was likely enriched predominantly by CCSNe during earlier epochs \citep{Beverage2025}.
These findings are in line with the general understanding at $z\sim0$.

Interestingly, this understanding has been challenged by an increasing number of \textit{JWST} results.
Some metal-poor galaxies at $z\gtrsim4$ are reported to have high C/O ratios \citep{DEugenio2024,Ji2025b,Nakajima2025,Scholtz2025b}, low Ne/O ratios \citep{Isobe2023c}, or high Ar/O ratios \citep{Bhattacharya2025}, attributed to the possible contributions of Population III (PopIII) SNe \citep[e.g.,][]{Vanni2024}, massive ($\gtrsim30$ $M_{\odot}$) CCSNe \citep{Watanabe2024}, and even theoretical pair-instability SNe (PISNe) whose progenitors have $\sim200$--300 $M_{\odot}$ \citep{Takahashi2018}, respectively.
Although \citet{Curti2025b} have recently reported that a Wolf-Rayet (WR) galaxy at $z\sim2$ has a detection of the [\feiii]$\lambda$4658 line, which has been commonly used at $z\sim0$ \citep[e.g.,][]{Izotov2006}, the [\feiii] line is usually very faint and hence extremely difficult to observe at higher $z$.
In a complementary manner, higher ionisation Fe lines \citep{Ji2024,Tacchella2025}, the Fe\,\textsc{ii} complex \citep{Ji2025d,Nakane2025}, and the ultra-violet (UV) stellar continuum shape \citep{Nakane2024,Nakane2025} have been utilised to estimate Fe/O ratios, some of which exceed the solar abundance ratio.
Such high Fe/O ratios in young galaxies may imply inclusions of (very) massive stars resulting in such as PISNe \citep[e.g.,][]{Isobe2022,Goswami2022,Fukushima2025} or even $>300\ M_{\odot}$ stars \citep{Kojima2021}, while the possibility of Type-Ia SNe with short delay times has also been discussed \citep{Nakane2025}.
This line of argument is similar to that for the \textit{JWST}‑discovered population of N/O‑enhanced galaxies (NOEGs; \citealt{Ji2025b}) at $z>4$ \citep[e.g.,][]{Bunker2023,Cameron2023,Isobe2023c,Topping2024a}, in that many papers have actively discussed the possibility of N-rich winds from WR stars \citep[e.g.,][]{Watanabe2024,Kobayashi2024,Fukushima2024}, very massive stars \citep{Vink2023}, and even supermassive stars with $10^{3}\ M_{\odot}$ \citep[e.g.,][]{Nagele2023,Charbonnel2023,Nandal2024a}, which are not necessarily required in some scenarios with AGB stars \citep[e.g.,][]{DAntona2023,Rizzuti2024,McClymont2025c}.

One of the key issues to be addressed here is whether these abundance ratios are common in $z>4$ galaxies, as abundance ratios based on fainter emission lines tend to be biased more strongly towards the readily observable population.
For example, the possibility of active galactic nuclei (AGNs) has been discussed for about half of the NOEGs reported so far \citep[e.g.,][]{Ubler2023,Ji2024,Maiolino2024a,Napolitano2024b}.
Complementary to individual galaxy studies, the stacking technique is commonly employed to characterise the typical properties of galaxy populations in which the target signals are not individually detectable \citep[e.g.,][]{Roberts-Borsani2024,Kumari2024,Glazer2025}.
In a previous work, we  focused on N/O ratios \citep{Isobe2025}, illustrating that the stacked spectrum of galaxies without reported AGNs \citep{Juodzbalis2025,Scholtz2025} has a lower N/O ratio than most of the NOEGs, implying their rarity.
Regarding other abundance ratios, several studies report low C/O ratios in stacked spectra of $z>4$ galaxies \citep{Hu2024,Hayes2025}, suggesting a gas composition dominated by CCSNe.
No other abundance ratios have been measured with stacked spectra at $z>4$.
In particular, compared to C/O and Ne/O ratios, the current small sample sizes of $z>4$ galaxies with Ar/O and Fe/O measurements prevent a reliable assessment of their general values.

Another important topic is the nature of surprisingly large dust reservoirs that have formed at high $z$ \citep[e.g.,][]{Watson2015,Dayal2022,Witstok2022,Witstok2023a,Algera2023,Algera2024,Sun2025}.
Dust grains are thought to consist mainly of carbonaceous and silicate grains containing C and Si, respectively \citep[e.g.,][]{Li2001}, and most interstellar Fe is thought to be incorporated into silicate grains \citep[e.g.][]{Zhukovska2018}.
Unlike carbonaceous grains, which arise mainly from low-mass AGB stars with 2--3 $M_{\odot}$, silicates are produced by massive AGB stars ($>4$--5 $M_{\odot}$) and CCSNe, indicating their much shorter enrichment delay time of $\lesssim30$--40 Myr \citep[e.g.,][]{Schneider2024}, and thus, their dominance within $\lesssim100$ Myr \citep[e.g.,][]{Hirashita2020}.
Simulations have predicted that silicate grains tend to be larger at earlier epochs of star formation \citep[e.g.,][]{Hou2017}, resulting in the flattening of the observed average dust attenuation curve at $z>4.5$ \citep{Markov2025,McKinney2025,Shivaei2025}.
Although direct observational evidence such as the UV bump for carbonaceous grains has been observed out to high redshifts \citep[e.g.,][]{Witstok2023b,Markov2025,Ormerod2025}, strong silicate features at restframe $\sim10$--20 $\mu$m \citep[e.g.,][]{Mathis1990,vanBreemen2011} are difficult to observe at high-$z$ ($z\sim2$--10), because they fall beyond the wavelength coverage of the \textit{JWST} Mid-Infrared Instrument (MIRI).
Alternatively, as for the Fe/O ratios at $z\sim0$ \citep[e.g.,][]{Izotov2006,Mendez-Delgado2024}, low Si/O ratios have been discussed to suggest dust depletion \citep[e.g.,][]{Garnett1995b}.
In addition, Si is a heavy $\alpha$-element produced by both CCSNe and Type-Ia SNe, suggesting a similar evolution to Ar enrichment.
Given that Ar is a noble gas with negligible dust depletion \citep{Sofia1998}, a significant deficit of Si compared to Ar might indicate Si depletion onto dust grains.

This paper aims at obtaining typical values of C/O, Ne/O, Ar/O, Si/O, and Fe/O ratios by stacking spectra of galaxies at $z>4$.
We carry out the stacking with different bins of galaxy properties to explore potential dependencies of the abundance ratios.
We present our data and sample in Section \ref{sec:datsamp}, analysis in Section \ref{sec:ana}, results and discussions in Section \ref{sec:res}, and conclusions in Section \ref{sec:con}.
Hereafter, we abbreviate
\oiii]$\lambda\lambda$1661,1666 to \oiii],
\ciii]$\lambda\lambda$1907,1909 to \ciii],
[\oii]$\lambda\lambda$3727,3729 to [\oii],
[\neiii]$\lambda$3869 to [\neiii],
[\feiii]$\lambda$4658 to [\feiii], and
[\ariii]$\lambda$7135 to [\ariii],
for simplicity.
Unless otherwise specified, abundance ratios hereafter simply refer to gas-phase ones.
Throughout this paper, we use the solar abundance ratios of \citet{Asplund2021}.
We assume a standard $\Lambda$CDM cosmology with parameters of $\Omega_{0}=0.315$ and $H_{0}=67.4\ \mathrm{km\ s^{-1}\ Mpc^{-1}}$ \citep{Planck2020}.

\section{Data and Sample} \label{sec:datsamp}
\begin{figure*}
	\centering
    \includegraphics[width=\textwidth]{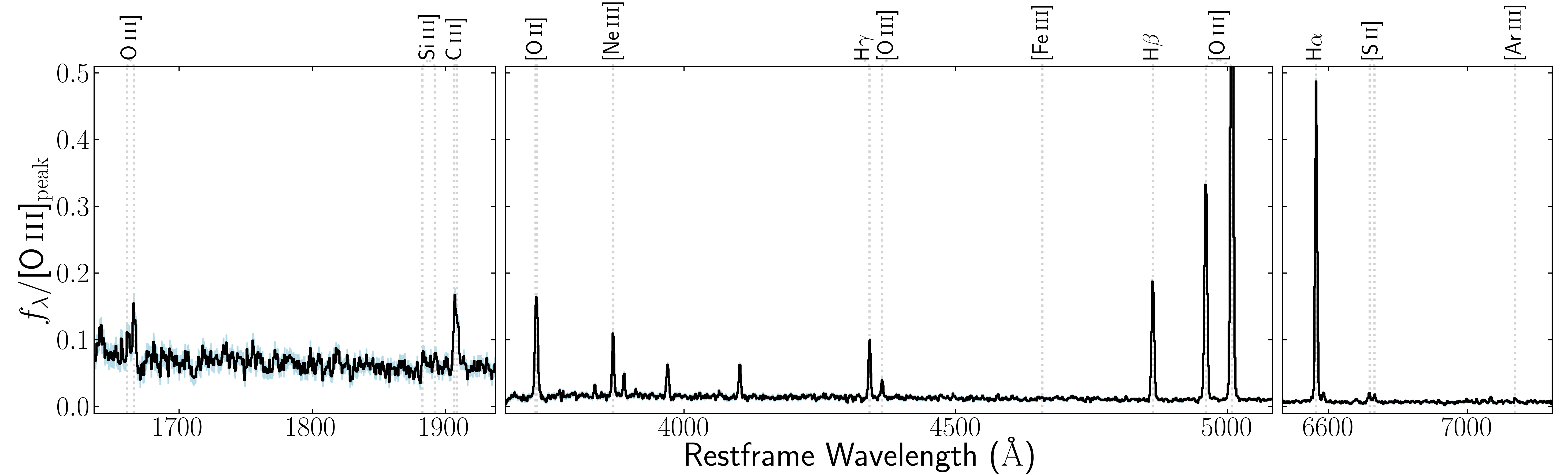}
    \caption{R1000 composite spectra of the all-sources stack (black) with the errors (lightblue), where the $y$ axis is in the unit of flux density per wavelength ($f_{\lambda}$) normalised by the peak of [\oiii]$\lambda$5007 ([\oiii]$_{\mathrm{peak}}$). The grey dotted lines with the black texts show the wavelengths of emission lines that are used to measure nebular properties in Section \ref{subsec:neb}.}
    \label{fig:stkspec}
\end{figure*}

\begin{table*}
	\centering
	\caption{Fundamental properties of our stacks. The value with the errors represents the median value with the range of the 16th and 84th percentiles of the sources used for each stack.}
	\label{tab:fund}
    \resizebox{\textwidth}{!}{
	\begin{tabular}{lccccccccccccc}
		\hline
        Property & All-sources & Low-$M_{*}$ & Mid-$M_{*}$ & High-$M_{*}$ & Low-SFR & Mid-SFR & High-SFR & Low-sSFR & Mid-sSFR & High-sSFR & Blue-\buv & Mid-\buv & Red-\buv\\
        \hline
        \# of spectra & 564 & 181 & 182 & 183 & 179 & 185 & 182 & 182 & 183 & 181 & 117 & 116 & 117\\
        $z$ & $5.27^{+0.99}_{-0.98}$ & $5.58^{+0.79}_{-1.13}$ & $5.27^{+1.11}_{-0.94}$ & $4.83^{+1.11}_{-0.70}$ & $5.26^{+0.71}_{-0.91}$ & $5.27^{+1.06}_{-1.09}$ & $5.37^{+0.95}_{-0.97}$ & $4.77^{+1.11}_{-0.64}$ & $5.26^{+0.80}_{-1.09}$ & $5.78^{+0.85}_{-0.97}$ & $5.86^{+0.76}_{-1.05}$ & $5.23^{+1.04}_{-0.86}$ & $4.87^{+1.04}_{-0.73}$\\
        $\log(M_{*}/M_{\odot})$ & $8.46^{+0.71}_{-0.71}$ & $7.76^{+0.29}_{-0.48}$ & $8.46^{+0.26}_{-0.20}$ & $9.14^{+0.45}_{-0.22}$ & $7.90^{+0.67}_{-0.62}$ & $8.45^{+0.60}_{-0.49}$ & $8.96^{+0.62}_{-0.55}$ & $9.05^{+0.50}_{-0.53}$ & $8.50^{+0.47}_{-0.49}$ & $7.96^{+0.45}_{-0.64}$ & $8.21^{+0.63}_{-0.39}$ & $8.60^{+0.47}_{-0.44}$ & $9.11^{+0.48}_{-0.42}$\\
        $\log(\mathrm{SFR}/M_{\odot}\,\mathrm{yr}^{-1})$ & $0.30^{+0.48}_{-0.52}$ & $-0.09^{+0.43}_{-0.47}$ & $0.30^{+0.37}_{-0.36}$ & $0.64^{+0.48}_{-0.41}$ & $-0.20^{+0.19}_{-0.40}$ & $0.30^{+0.13}_{-0.14}$ & $0.76^{+0.39}_{-0.19}$ & $0.28^{+0.49}_{-0.63}$ & $0.30^{+0.48}_{-0.47}$ & $0.30^{+0.45}_{-0.51}$ & $0.29^{+0.37}_{-0.34}$ & $0.41^{+0.33}_{-0.35}$ & $0.68^{+0.49}_{-0.51}$\\
        $\log(\mathrm{sSFR/Gyr^{-1}})$ & $0.81^{+0.55}_{-0.54}$ & $1.26^{+0.27}_{-0.44}$ & $0.82^{+0.46}_{-0.30}$ & $0.42^{+0.35}_{-0.45}$ & $0.82^{+0.53}_{-0.71}$ & $0.82^{+0.53}_{-0.55}$ & $0.79^{+0.57}_{-0.38}$ & $0.29^{+0.25}_{-0.41}$ & $0.81^{+0.14}_{-0.16}$ & $1.35^{+0.25}_{-0.21}$ & $1.07^{+0.36}_{-0.43}$ & $0.74^{+0.55}_{-0.49}$ & $0.57^{+0.36}_{-0.53}$\\
        \buv & $-2.13^{+0.44}_{-0.37}$ & $-2.43^{+0.32}_{-0.23}$ & $-2.17^{+0.32}_{-0.27}$ & $-1.88^{+0.43}_{-0.37}$ & $-2.24^{+0.40}_{-0.40}$ & $-2.17^{+0.32}_{-0.35}$ & $-1.98^{+0.49}_{-0.38}$ & $-1.94^{+0.44}_{-0.32}$ & $-2.16^{+0.40}_{-0.28}$ & $-2.34^{+0.36}_{-0.27}$ & $-2.49^{+0.16}_{-0.18}$ & $-2.13^{+0.13}_{-0.09}$ & $-1.71^{+0.33}_{-0.16}$\\
        \hline
	\end{tabular}
    }
\end{table*}

We analyse the same spectroscopic dataset as \citet{Isobe2025}, which use the \textit{JWST} Advanced Deep Extragalactic Survey (JADES; \citealt{Eisenstein2023}) data obtained with the NIRSpec micro-shutter array (MSA; \citealt{Jakobsen2022}; \citealt{Ferruit2022}) in the GOODS-S and GOODS-N fields.
The dataset contains the complete observations of the four programmes: PIDs 1180, 1181, 1210, and 3215 \citep{Bunker2024,DEugenio2025,Eisenstein2023b} and part of the two programmes: PIDs 1286 and 1287.
JADES provides medium-resolution grating and low-resolution prism observations with resolution $R\sim1000$ and $R\sim100$ (R1000 and R100, hereafter).
This paper uses the R1000 data to resolve important emission line pairs separated by a small wavelength (Section \ref{subsec:emis} for more details).
The R1000 data were obtained with three bands of F070LP-G140M, F170LP-G235M, and F290LP-G395M, which cover a wide wavelength range of 1--5.3 $\mu$m.

A description of the JADES data reduction is available in the NIRSpec Data Release 3 \citep[DR3;][]{DEugenio2025} and the DR4 papers \citep{Scholtz2025_DR4}, so only the main points are summarised here.
The NIRSpec Guaranteed Time Observations (GTO) Team reduced the observed data using the pipeline constructed by the European Space Agency (ESA) NIRSpec Science Operations Team \citep{Ferruit2022} and the NIRSpec GTO Team \citep{Oliveira2018}.
Instead of the standard 5-pixel (0.5'') full-microshutter extraction aperture \citep{DEugenio2025}, we choose a 3-pixel box-car aperture to increase the signal-to-noise ratio (S/N) of the spectrum, particularly for compact sources, at shorter wavelengths with the narrower \textit{JWST} point spread functions.
The 3-pixel box-car aperture extractions are available in the DR4 \citep{Scholtz2025_DR4}.
The standard pipeline incorporates error propagation and employs variance-conserving resampling to conservatively account for correlated noise \citep{Dorner2016}.

We construct our parent sample with the following criteria:
\begin{itemize}
    \item Spectra with flags of `6', `7', or `8' \citep{DEugenio2025}, whose $z$ values are reliably determined with multiple emission lines
    \item Spectra with [\oiii]$\lambda$5007 whose S/N ratio is higher than 3 based on the R1000 observations (Section \ref{subsec:stk})
    \item Spectra with $z=4$--7 to widely cover from \oiii] to [\ariii], where the extensive coverage provided by all three NIRSpec bands means that most of the rest-frame wavelength range is covered for each galaxy (see Section \ref{subsec:emis})
    \item Excluding broad-line active galactic nuclei (AGNs) based on H$\alpha$ line broadening \citep{Juodzbalis2025} and AGN candidates selected with emission line diagnostics \citep{Scholtz2025}.
\end{itemize}
These selection criteria provide 564 spectra.
Note that the last criterion is set to mitigate the possibility of anomalous chemical evolution originating from AGN activity.
In fact, the stack of similarly selected spectra (i.e., `Non-AGN' stack in \citealt{Isobe2025}) does not exhibit N/O and N/C ratios as high as those of high-$z$ NOEGs, but the broad-line AGN stack does.
However, it is worth noting that our sample does not fully exclude a hidden population of low-luminosity AGNs whose signatures are not identified individually \citep{Geris2025}.

The majority of our sample has measurements of the following galaxy properties: stellar mass ($M_{*}$), star-formation rate (SFR), specific SFR ($\mathrm{sSFR}\equiv\mathrm{SFR}/M_{*}$), and UV continuum slope \buv, which allow us to investigate possible dependencies of chemical abundances on these properties (see Section \ref{subsec:stk}).
Most of our spectra have the corresponding NIRCam imaging from the JADES \citep{Rieke2023,Eisenstein2023b}, the \textit{JWST} Extragalactic Medium-band Survey (JEMS; \citealt{Williams2023}), and the First Reionisation Epoch Spectroscopic Complete Survey (FRESCO; \citealt{Oesch2023}), whose source detection and photometric measurements have been conducted by Robertson et al. (in preparation).
We follow the spectral energy distribution (SED) fitting routine of \citet{Simmonds2025}, except that the redshift is fixed to the value from the NIRSpec data.
Using Kron photometry convolved to a common resolution with an error floor of 5\% on each band, we derive $M_{*}$ and SFR values using the SED fitting code \texttt{Prospector} \citep[][]{Johnson2019,Johnson2021}, following the setup of \citealt{Tacchella2022} with a continuity star-formation history (SFH) from \citet{Leja2019}, which comprises eight SFR bins.
The shortest time bin is 5 Myr, while the rest of the bins are equally spaced in log space depending on the redshift.
The ratios between adjacent SFR bins are allowed to vary following a Student's t-distribution \citep{Student1908} with a width of 0.3.
Although the stellar population used for the SED fitting is not identical to that adopted for the ionisation corrections (Section \ref{subsec:neb}), we confirm that the SED fitting results vary only within the quoted uncertainties under standard IMF assumptions and when binaries are included, based on tests using a subset of similar photometric data.

In this work, we use the SFR values averaged over the recent 10 Myr (so-called SFR$_{10}$).
We have 546 spectra with measurements of $M_{*}$ and SFR$_{10}$, which comprise the bulk of our 564 spectra.
The remaining 18 sources without $M_{*}$ or SFR$_{10}$ do not have enough NIRCam photometric points to constrain the best-fit SED model.
We derive sSFR values from these $M_{*}$ and SFR$_{10}$ values.
We use \buv\ values derived by \citet{Saxena2024} from the restframe 1340--2700 \AA\ continuum of the JADES R100 spectra, which requires $S/N>3$ in the restframe UV continuum.
We have 350 of our 564 spectra that satisfy the criterion and have the \buv\ values.
Note that \citet{Saxena2024} have reported sources at $z>5.5$, while their \buv\ measurements exist at $z>4$, which we use in this paper.
It should also be mentioned that the $S/N$ cut bias the sample towards higher $M_{*}$ galaxies.

\section{Analysis} \label{sec:ana}
\begin{figure*}
	\includegraphics[width=\textwidth]{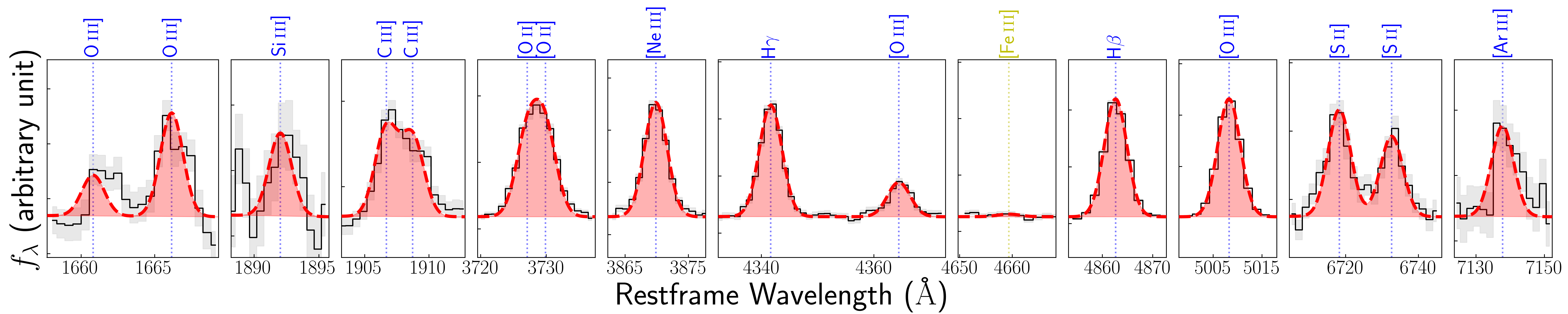}
    \caption{Best-fit model (red) of the all-sources stack (black) with its errors (grey). The $y$ axis in each panel is rescaled for visualisation purpose, while the scale for the panel of [\feiii] is the same as that of H$\gamma$+[\oiii]$\lambda$4363. The blue texts indicate detected emission lines, while the yellow text corresponds to the undetected line. All of the emission lines shown here are detected, except for [\feiii].}
    \label{fig:bestfit}
\end{figure*}

\begin{table*}
	\centering
	\caption{Emission line fluxes of our stacks normalised by H$\beta$ before dust correction. The upper limits are $3\sigma$. $^{\dagger}$: Doublets that are not fully separated in the R1000 spectra.}
	\label{tab:emis}
    \resizebox{\textwidth}{!}{
	\begin{tabular}{lccccccccccccc}
		\hline
        Property & All-sources & Low-$M_{*}$ & Mid-$M_{*}$ & High-$M_{*}$ & Low-SFR & Mid-SFR & High-SFR & Low-sSFR & Mid-sSFR & High-sSFR & Blue-\buv & Mid-\buv & Red-\buv\\
        \hline
        \oiii] & $15.6\pm1.8$ & $27.2\pm3.0$ & $13.1\pm2.9$ & $<8.0$ & $26.4\pm5.3$ & $17.3\pm2.6$ & $13.1\pm2.1$ & $13.7\pm4.3$ & $10.2\pm3.1$ & $20.5\pm2.1$ & $24.8\pm2.6$ & $23.9\pm2.9$ & $<8.6$\\
        \siiii]$\lambda$1892 & $4.8\pm1.3$ & $<8.5$ & $8.1\pm1.9$ & $<6.6$ & $<11.9$ & $8.9\pm2.1$ & $<4.6$ & $<8.6$ & $7.7\pm2.3$ & $<4.8$ & $<6.4$ & $7.2\pm2.3$ & $<6.0$\\
        \ciii]$\lambda$1907$^{\dagger}$ & $19.5\pm1.5$ & $27.2\pm2.9$ & $22.8\pm2.1$ & $10.8\pm2.1$ & $26.7\pm4.2$ & $26.7\pm2.3$ & $14.0\pm1.6$ & $12.3\pm2.6$ & $20.1\pm2.6$ & $20.5\pm1.8$ & $23.0\pm2.5$ & $25.0\pm2.3$ & $17.2\pm2.3$\\
        \ciii]$\lambda$1909$^{\dagger}$ & $17.5\pm1.7$ & $19.0\pm2.6$ & $18.3\pm2.3$ & $13.6\pm2.5$ & $23.4\pm4.4$ & $19.1\pm2.4$ & $14.8\pm1.6$ & $18.6\pm3.1$ & $17.3\pm2.6$ & $17.0\pm1.6$ & $20.0\pm2.3$ & $14.7\pm2.4$ & $14.1\pm2.4$\\
        
        [\oii]$\lambda$3726$^{\dagger}$ & $44.1\pm1.8$ & $20.1\pm2.4$ & $43.4\pm2.5$ & $66.6\pm3.3$ & $26.8\pm3.8$ & $37.6\pm2.4$ & $55.6\pm2.2$ & $58.3\pm3.2$ & $50.8\pm2.2$ & $21.5\pm1.6$ & $23.2\pm2.3$ & $46.5\pm2.5$ & $64.1\pm3.5$\\
        
        [\oii]$\lambda$3729$^{\dagger}$ & $46.0\pm1.9$ & $19.9\pm2.3$ & $43.5\pm2.4$ & $88.3\pm3.5$ & $24.5\pm3.4$ & $46.7\pm2.5$ & $64.6\pm2.4$ & $81.8\pm3.6$ & $50.1\pm2.4$ & $26.7\pm1.6$ & $27.3\pm2.1$ & $54.3\pm2.8$ & $80.2\pm3.7$\\
        
        [\neiii] & $42.0\pm1.0$ & $36.1\pm1.8$ & $44.9\pm1.4$ & $43.8\pm1.5$ & $38.4\pm2.7$ & $43.9\pm1.4$ & $43.4\pm1.1$ & $42.6\pm1.9$ & $42.1\pm1.5$ & $39.4\pm1.2$ & $41.3\pm1.5$ & $46.7\pm1.6$ & $43.7\pm1.5$\\
        H$\gamma$ & $42.2\pm0.9$ & $43.0\pm1.8$ & $45.4\pm1.3$ & $39.4\pm1.4$ & $41.0\pm2.3$ & $44.7\pm1.4$ & $42.5\pm1.1$ & $39.8\pm1.8$ & $43.3\pm1.3$ & $44.1\pm1.3$ & $42.6\pm1.7$ & $44.1\pm1.5$ & $40.3\pm1.3$\\
        
        [\oiii]$\lambda$4363 & $13.1\pm0.8$ & $15.2\pm1.4$ & $14.4\pm1.1$ & $9.7\pm1.3$ & $14.5\pm2.0$ & $16.2\pm1.2$ & $9.4\pm0.8$ & $12.3\pm1.6$ & $11.7\pm1.2$ & $12.9\pm1.1$ & $16.2\pm1.2$ & $13.7\pm1.2$ & $11.7\pm1.2$\\
        
        [\feiii] & $<2.3$ & $<4.0$ & $<3.8$ & $<3.3$ & $<5.9$ & $<3.6$ & $<2.7$ & $<3.9$ & $<3.5$ & $3.0\pm0.9$ & $4.7\pm1.1$ & $<3.6$ & $<3.4$\\
        H$\beta$ & $100.0\pm1.3$ & $100.0\pm2.0$ & $100.0\pm1.8$ & $100.0\pm2.1$ & $100.0\pm2.9$ & $100.0\pm1.6$ & $100.0\pm1.5$ & $100.0\pm2.0$ & $100.0\pm1.8$ & $100.0\pm1.5$ & $100.0\pm1.6$ & $100.0\pm2.1$ & $100.0\pm1.9$\\
        
        [\oiii]$\lambda$5007 & $569.8\pm2.5$ & $486.8\pm3.3$ & $632.5\pm3.5$ & $612.8\pm5.5$ & $492.1\pm4.1$ & $571.0\pm3.3$ & $620.5\pm3.1$ & $590.4\pm5.2$ & $595.1\pm3.6$ & $561.7\pm3.1$ & $555.1\pm2.9$ & $609.5\pm4.2$ & $623.1\pm4.6$\\
        H$\alpha$ & $338.9\pm2.1$ & $323.3\pm3.3$ & $350.7\pm2.7$ & $362.7\pm3.5$ & $321.3\pm3.7$ & $335.9\pm2.5$ & $350.3\pm2.8$ & $336.8\pm3.6$ & $346.2\pm3.0$ & $336.5\pm2.5$ & $328.4\pm2.8$ & $342.5\pm2.8$ & $356.0\pm4.0$\\
        
        [\sii]$\lambda$6716 & $10.4\pm0.8$ & $5.4\pm1.4$ & $9.3\pm1.1$ & $21.0\pm1.4$ & $6.5\pm2.0$ & $8.8\pm1.2$ & $14.2\pm1.2$ & $20.0\pm1.7$ & $9.7\pm1.3$ & $6.4\pm1.1$ & $5.1\pm1.3$ & $11.3\pm1.2$ & $15.5\pm1.3$\\
        
        [\sii]$\lambda$6731 & $7.9\pm0.7$ & $<3.9$ & $6.4\pm1.1$ & $13.9\pm1.4$ & $<6.0$ & $5.1\pm1.0$ & $11.2\pm1.1$ & $11.5\pm1.6$ & $7.8\pm1.2$ & $4.7\pm0.9$ & $<3.9$ & $8.0\pm1.2$ & $11.1\pm1.2$\\

        [\ariii] & $4.8\pm0.8$ & $<4.8$ & $4.6\pm1.2$ & $6.9\pm1.5$ & $<6.4$ & $5.5\pm1.1$ & $6.2\pm1.0$ & $6.8\pm1.5$ & $5.7\pm1.2$ & $<3.6$ & $<4.4$ & $5.8\pm1.3$ & $6.3\pm1.2$\\
        \hline
	\end{tabular}
    }
\end{table*}

\begin{figure}
	\centering
    \includegraphics[width=\columnwidth]{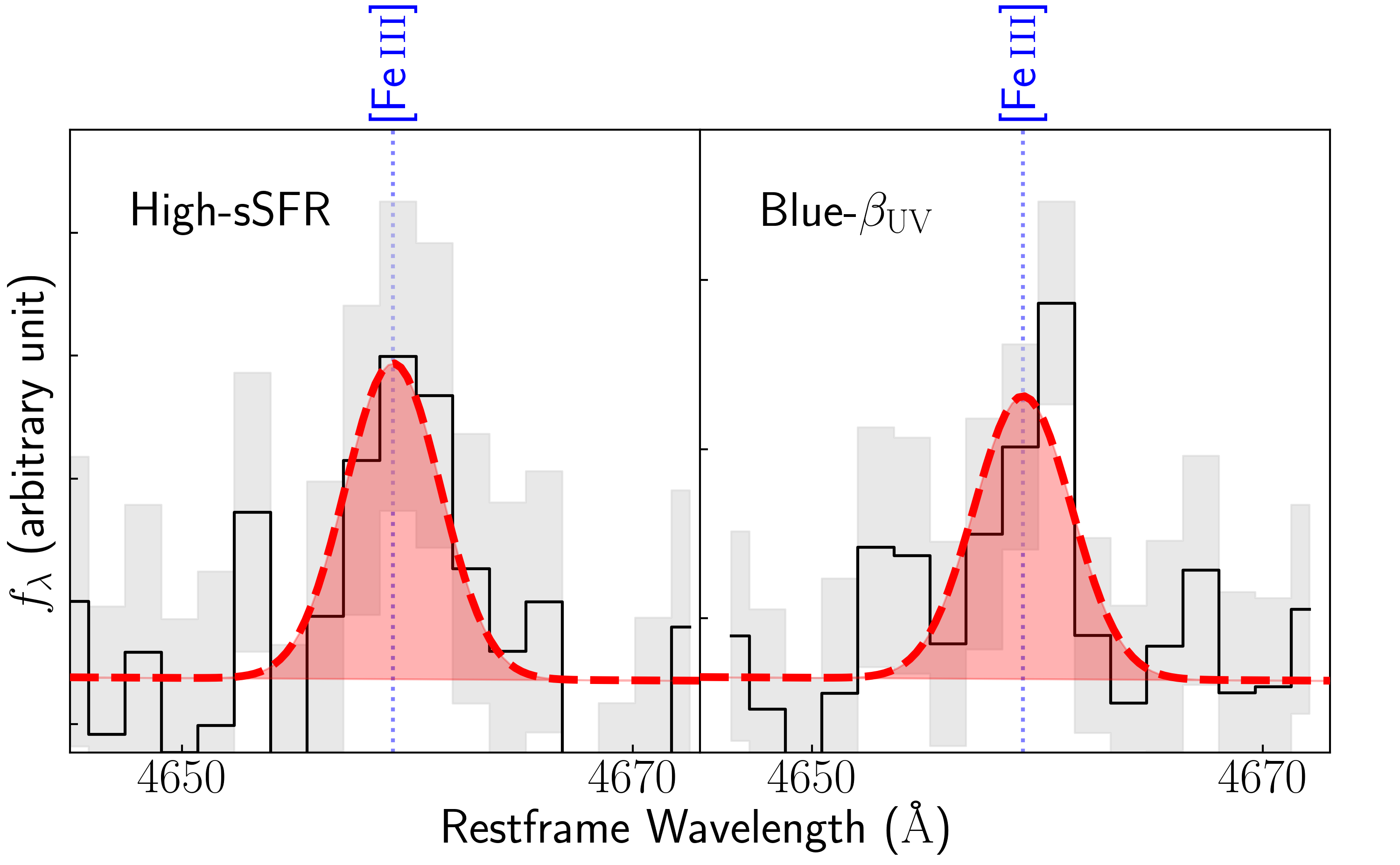}
    \caption{Same as Figure \ref{fig:bestfit} but it highlights spectra of the high-sSFR and blue-\buv\ stacks around [\feiii]. These stacks are the only spectra
    whose $S/N$ ratio of [\feiii] exceeds 3.}
    \label{fig:fe3}
\end{figure}

\begin{table*}
	\centering
	\caption{Nebular properties of our stacks. $^{\dagger}$: 1$\sigma$ upper limit because the measured value reaches the low density limit. The other limits are 3$\sigma$.}
	\label{tab:neb}
    \resizebox{\textwidth}{!}{
	\begin{tabular}{lccccccccccccc}
		\hline
        Property & All-sources & Low-$M_{*}$ & Mid-$M_{*}$ & High-$M_{*}$ & Low-SFR & Mid-SFR & High-SFR & Low-sSFR & Mid-sSFR & High-sSFR & Blue-\buv & Mid-\buv & Red-\buv\\
        \hline
        $E(B-V)$ & $0.18\pm0.01$ & $0.15\pm0.01$ & $0.18\pm0.02$ & $0.25\pm0.02$ & $0.16\pm0.02$ & $0.16\pm0.01$ & $0.20\pm0.00$ & $0.19\pm0.03$ & $0.19\pm0.00$ & $0.16\pm0.00$ & $0.16\pm0.01$ & $0.18\pm0.00$ & $0.23\pm0.02$\\
        $T_{\mathrm{e}}$[\oiii]\,($10^{4}$\,K) & $1.72^{+0.05}_{-0.05}$ & $2.02^{+0.12}_{-0.11}$ & $1.71^{+0.07}_{-0.07}$ & $1.46^{+0.09}_{-0.09}$ & $1.96^{+0.16}_{-0.16}$ & $1.92^{+0.08}_{-0.08}$ & $1.42^{+0.05}_{-0.05}$ & $1.64^{+0.11}_{-0.11}$ & $1.60^{+0.08}_{-0.08}$ & $1.71^{+0.08}_{-0.08}$ & $1.95^{+0.09}_{-0.09}$ & $1.70^{+0.08}_{-0.08}$ & $1.57^{+0.08}_{-0.08}$\\
        $\log(n_{\mathrm{e}}[\text{\textsc{S\,ii}}]/\mathrm{cm^{-3}})$ & $2.10^{+0.51}_{-1.50}$ & $<1.10^{\dagger}$ & $<2.66^{\dagger}$ & $<2.11^{\dagger}$ & $<2.34^{\dagger}$ & $<2.25^{\dagger}$ & $2.30^{+0.39}_{-1.69}$ & $<0.60^{\dagger}$ & $2.36^{+0.57}_{-1.76}$ & $1.94^{+1.01}_{-1.34}$ & $<1.70^{\dagger}$ & $1.31^{+1.34}_{-0.71}$ & $1.61^{+0.89}_{-1.01}$\\
        \met & $7.71^{+0.01}_{-0.01}$ & $7.45^{+0.01}_{-0.01}$ & $7.75^{+0.01}_{-0.01}$ & $7.96^{+0.01}_{-0.01}$ & $7.50^{+0.01}_{-0.02}$ & $7.60^{+0.01}_{-0.01}$ & $7.97^{+0.01}_{-0.01}$ & $7.81^{+0.01}_{-0.01}$ & $7.81^{+0.01}_{-0.01}$ & $7.67^{+0.01}_{-0.01}$ & $7.55^{+0.01}_{-0.01}$ & $7.75^{+0.01}_{-0.01}$ & $7.88^{+0.01}_{-0.01}$\\
        ICF($\mathrm{C^{2+}/O^{2+}}$) & $0.95^{+0.00}_{-0.00}$ & $1.03^{+0.01}_{-0.01}$ & $0.96^{+0.00}_{-0.00}$ & $0.87^{+0.01}_{-0.00}$ & $0.99^{+0.01}_{-0.01}$ & $0.96^{+0.00}_{-0.00}$ & $0.92^{+0.00}_{-0.01}$ & $0.88^{+0.01}_{-0.00}$ & $0.94^{+0.00}_{-0.00}$ & $1.02^{+0.01}_{-0.00}$ & $1.01^{+0.01}_{-0.01}$ & $0.95^{+0.00}_{-0.00}$ & $0.88^{+0.01}_{-0.01}$\\
        $\log(\mathrm{C/O})$ & $-0.93^{+0.06}_{-0.05}$ & $-0.99^{+0.06}_{-0.06}$ & $-0.81^{+0.12}_{-0.09}$ & $>-0.94$ & $-0.96^{+0.11}_{-0.09}$ & $-0.85^{+0.07}_{-0.07}$ & $-1.05^{+0.09}_{-0.07}$ & $-1.00^{+0.16}_{-0.14}$ & $-0.77^{+0.15}_{-0.12}$ & $-1.01^{+0.05}_{-0.05}$ & $-1.00^{+0.06}_{-0.06}$ & $-1.09^{+0.07}_{-0.06}$ & $>-0.82$\\
        ICF($\mathrm{Ne^{2+}/O^{2+}}$) & $0.87^{+0.00}_{-0.00}$ & $0.92^{+0.00}_{-0.01}$ & $0.88^{+0.00}_{-0.00}$ & $0.81^{+0.01}_{-0.00}$ & $0.90^{+0.01}_{-0.01}$ & $0.88^{+0.00}_{-0.00}$ & $0.86^{+0.00}_{-0.01}$ & $0.81^{+0.01}_{-0.00}$ & $0.87^{+0.01}_{-0.00}$ & $0.93^{+0.00}_{-0.00}$ & $0.91^{+0.01}_{-0.01}$ & $0.87^{+0.00}_{-0.00}$ & $0.82^{+0.01}_{-0.01}$\\
        $\log(\mathrm{Ne/O})$ & $-0.72^{+0.01}_{-0.01}$ & $-0.74^{+0.02}_{-0.02}$ & $-0.74^{+0.02}_{-0.01}$ & $-0.72^{+0.02}_{-0.01}$ & $-0.72^{+0.03}_{-0.04}$ & $-0.73^{+0.01}_{-0.01}$ & $-0.71^{+0.01}_{-0.01}$ & $-0.76^{+0.02}_{-0.02}$ & $-0.73^{+0.02}_{-0.01}$ & $-0.73^{+0.01}_{-0.01}$ & $-0.73^{+0.02}_{-0.02}$ & $-0.71^{+0.02}_{-0.02}$ & $-0.74^{+0.02}_{-0.01}$\\
        ICF($\mathrm{Ar^{2+}/O^{2+}}$) & $0.91^{+0.01}_{-0.01}$ & $1.12^{+0.03}_{-0.03}$ & $0.95^{+0.01}_{-0.01}$ & $0.77^{+0.01}_{-0.01}$ & $1.03^{+0.04}_{-0.03}$ & $0.94^{+0.01}_{-0.01}$ & $0.87^{+0.01}_{-0.01}$ & $0.80^{+0.01}_{-0.01}$ & $0.90^{+0.01}_{-0.01}$ & $1.11^{+0.02}_{-0.02}$ & $1.08^{+0.02}_{-0.02}$ & $0.91^{+0.01}_{-0.01}$ & $0.80^{+0.01}_{-0.01}$\\
        $\log(\mathrm{Ar/O})$ & $-2.59^{+0.07}_{-0.08}$ & $<-2.37$ & $-2.64^{+0.11}_{-0.13}$ & $-2.64^{+0.08}_{-0.10}$ & $<-2.30$ & $-2.48^{+0.08}_{-0.09}$ & $-2.62^{+0.06}_{-0.08}$ & $-2.54^{+0.09}_{-0.10}$ & $-2.58^{+0.09}_{-0.10}$ & $<-2.62$ & $<-2.50$ & $-2.54^{+0.09}_{-0.10}$ & $-2.63^{+0.08}_{-0.09}$\\
        ICF($\mathrm{Si^{2+}/O^{2+}}$) & $1.16^{+0.01}_{-0.01}$ & $1.31^{+0.03}_{-0.03}$ & $1.19^{+0.01}_{-0.01}$ & $1.05^{+0.01}_{-0.01}$ & $1.24^{+0.03}_{-0.02}$ & $1.18^{+0.01}_{-0.01}$ & $1.14^{+0.01}_{-0.01}$ & $1.05^{+0.03}_{-0.00}$ & $1.14^{+0.04}_{-0.00}$ & $1.33^{+0.02}_{-0.01}$ & $1.27^{+0.02}_{-0.02}$ & $1.15^{+0.01}_{-0.01}$ & $1.08^{+0.01}_{-0.01}$\\
        $\log(\mathrm{Si/O})$ & $-1.81^{+0.11}_{-0.15}$ & $<-1.67$ & $-1.49^{+0.14}_{-0.14}$ & $\cdots$ & $<-1.55$ & $-1.52^{+0.12}_{-0.14}$ & $<-1.87$ & $<-1.56$ & $-1.46^{+0.20}_{-0.18}$ & $<-1.86$ & $<-1.78$ & $-1.82^{+0.13}_{-0.18}$ & $\cdots$\\
        ICF($\mathrm{Fe^{2+}/O^{+}}$) & $0.72^{+0.00}_{-0.00}$ & $0.68^{+0.01}_{-0.00}$ & $0.71^{+0.00}_{-0.00}$ & $0.77^{+0.00}_{-0.00}$ & $0.69^{+0.00}_{-0.01}$ & $0.71^{+0.00}_{-0.00}$ & $0.73^{+0.00}_{-0.00}$ & $0.75^{+0.00}_{-0.00}$ & $0.72^{+0.01}_{-0.00}$ & $0.68^{+0.00}_{-0.00}$ & $0.69^{+0.00}_{-0.00}$ & $0.72^{+0.00}_{-0.00}$ & $0.76^{+0.00}_{-0.00}$\\
        $\log(\mathrm{Fe/O})$ & $<-1.56$ & $<-0.89$ & $<-1.25$ & $<-1.59$ & $<-0.97$ & $<-1.22$ & $<-1.68$ & $<-1.43$ & $<-1.48$ & $-1.17^{+0.11}_{-0.22}$ & $-0.98^{+0.16}_{-0.05}$ & $<-1.35$ & $<-1.56$\\
        \hline
	\end{tabular}
    }
\end{table*}

\subsection{Spectral stacking in different galaxy property bins} \label{subsec:stk}

We produce stacked spectra of the R1000 data in the same manner as \citet{Isobe2025} and \citet{Geris2025}, as summarised below.
We combine spectra of different NIRSpec bands for a given source.
When the wavelengths of two NIRSpec bands overlap, we adopt the band with the longer wavelength, as it is generally more sensitive \citep{Ferruit2022}.
Conducting emission line fitting for individual spectra, we obtain redshifts and [\oiii]$\lambda$5007 fluxes.
This line fitting provides redshifts, [\oiii]$\lambda$5007 fluxes, and S/N ratios, which are used for the sample selection (Section \ref{sec:datsamp}).
We shift the individual spectra to the restframe based on the $z$ values, with resampling onto a common wavelength grid.
We set a spectral pixel size to half the full-width half maximum of the NIRSpec line-spread function (LSF) based on point-like sources derived by \citet{deGraaff2024}.
This LSF has a spectral resolution typically twice as high as the nominal resolution assumed for uniformly illuminated shutters.
We resample the spectra using \texttt{spectres} \citep{Carnall2017}, which propagates the errors associated with the individual spectra.

Finally, we renormalise the resampled spectra by their [\oiii]$\lambda$5007 fluxes and construct the stacked spectrum by taking the median of the normalised spectra at each wavelength of the common wavelength grid, without applying any weighting.
This forms the basis of all results presented in this paper.
This stacking approach reduces the impact of a few bright outliers, allowing the resulting spectrum to more faithfully reflect the typical properties of the individual galaxies.
The statistical meaning of our stacking procedure is discussed in Appendix \ref{apsec:stat}.

As the uncertainty of the median cannot generally be derived analytically, we perform Monte Carlo simulations to estimate errors of the median stacked spectra.
We generate 1000 stacked spectra by randomly perturbing the individual spectra according to their errors, assuming a normal distribution, and then compute the standard deviation at each wavelength.
In the case of ``mean'' stacking, we have verified that this method reproduces analytic error propagation of the individual spectra \citep[also][]{Isobe2025}.
We do not use bootstrap errors for emission lines, as they tend to overestimate the flux variance by incorporating significant variations in the underlying continuum emission across different sources.

Following the method presented above, we conduct spectral stacking for all of our 564 spectra, which is referred to as the ``all-sources stack'' hereafter, shown in Fig.~\ref{fig:stkspec}.
In addition, we divide our parent sample into thirds based on $M_{*}$, SFR, sSFR, and \buv\ values (Section \ref{sec:datsamp}).
We define the low-value stack as comprising sources with the given property below the 33rd percentile, the mid-value stack as those between the 33rd and 67th percentiles, and the high-value stack as those above the 67th percentile.
This is made for statistical purposes only to ensure enough sources in each bin.
When referring to individual properties $M_{*}$, SFR, and sSFR, we replace `value' with the corresponding property (e.g., low-$M_{*}$ stack).
However, we specially call the low-, mid-, and high-value stacks of \buv\ as blue-, mid-, red-\buv\ stacks, respectively, because this terminology is widely used and intuitively understood.
The different stacks are shown in Fig.~\ref{apfig:stkspec}, while the number of stacked spectra and median properties of each stack are listed in Table \ref{tab:fund}.
We stress that more than 100 spectra are used for every stack.

\subsection{Emission line measurements} \label{subsec:emis}
We model our stacked spectra in the wide wavelength range of $\simeq1500$--7300 \AA, which is set to be covered by more than 70\% of the spectra in each stack.
We assume that each emission line profile has a velocity width $\sigma=\sqrt{\sigma_{\mathrm{int}}^{2}+\sigma_{\mathrm{LSF}}^{2}}$, where $\sigma_{\mathrm{int}}$ is an intrinsic velocity width common to all emission lines, and $\sigma_{\mathrm{LSF}}$ is the velocity width of \citet{deGraaff2024}'s LSF.
Given a potential Balmer break or jump, continua with $\lambda<3640$ and $>3820$ \AA\ are modelled with independent power-law functions, which are connected with a linear function within $3640\leq\lambda\leq3820$ \AA.
We perform $\chi^{2}$ fitting to our stacked spectra.
We define a line with $S/N>3$ as a detection, and adopt $3\sigma$ upper limits for undetected lines.
The best-fit model of the all-sources stack is shown in Figure \ref{fig:bestfit} as an example.

Table \ref{tab:emis} summarises the emission line fluxes used to measure nebular properties in Section \ref{subsec:neb}.
Every stack has a detection of [\oiii]$\lambda$4363, which provides a constraint on the electron temperature $T_{\mathrm{e}}$, ensuring that chemical abundances can be measured with the direct-$T_{\mathrm{e}}$ method (e.g., \citealt{Peimbert1967}; Section \ref{subsec:neb}).
We find [\neiii] and \ciii] detections in all our stacks, many of which have [\ariii] detections.

Interestingly, five of our stacks (the all-sources and all mid-value stacks) have \siiii]$\lambda$1892 detections, which are difficult to observe even at $z\sim0$ \citep[e.g.,][]{Berg2019}.
This allows us to obtain Si/O ratios based on the direct-$T_{\mathrm{e}}$ method at $z>4$.
We note that we do not use \siiii]$\lambda$1883 fluxes to measure Si/O ratios due to the nearby absorption feature at $\sim1881$ \AA\ seen in most of our stacks.
Additionally, we detect the [\feiii] line in two of our stacks (the high-sSFR and blue-\buv\ stacks) as shown in Figure \ref{fig:fe3}, which is widely used to determine Fe/O ratios at $z\sim0$ \citep[e.g.,][]{Izotov2006,Mendez-Delgado2024}.

\subsection{Nebular properties} \label{subsec:neb}

We obtain the nebular colour excess $E(B-V)$, electron temperature of [\oiii] ($T_{\mathrm{e}}$[\oiii]), and electron density of [\sii] ($n_{\mathrm{e}}$[\sii]) values of our stacks.
We use PyNeb \citep{Luridiana2015} to obtain the $E(B-V)$ value from the combination of observed H$\beta$/H$\alpha$, H$\gamma$/H$\alpha$, and H$\gamma$/H$\beta$ ratios by assuming Case B recombination, the $T_{\mathrm{e}}$[\oiii] value from the [\oiii]$\lambda$4363/[\oiii]$\lambda$5007 ratio, and the $n_{\mathrm{e}}$[\sii] value from the [\sii]$\lambda$6716/[\sii]$\lambda$6731 ratio.
Appendix \ref{apsec:atom} reports the atomic data that we use in this paper.
We calculate $E(B-V)$, $T_{\mathrm{e}}$[\oiii], and $n_{\mathrm{e}}$[\sii] values iteratively so that their values are consistent with each other \citep{Isobe2022}.
Note that, when we adopt the attenuation curve of \citet{Calzetti2000}, our stacks have $E(B-V)=0.15$--0.25, which corresponds to the $V$-band attenuation $A_{V}=0.35$--0.58.
\citet{Shivaei2025} have recently reported that the attenuation curve of galaxies with attenuation in the range of $0.1<A_{V}<0.6$ at $z=3$--7 lies between the attenuation curve of \citet{Calzetti2000} and the extinction curve of the Small Magellanic Cloud (SMC; \citealt{Gordon2003}), which suggests that either of the curves is applicable for our stacks.
We adopt the \citet{Calzetti2000}'s curve in this paper, as it provides abundance ratios that are more conservative for our conclusions than those of the SMC, as mentioned later in this section.
We derive $n_{\mathrm{e}}$[\sii] in the range $\log(n_{\mathrm{e}}[\text{\textsc{S\,ii}}]/\mathrm{cm^{-3}})=0.6$--4.7.
We estimate the errors of these properties using the 1$\sigma$ errors of the emission line ratios.
With the obtained $E(B-V)$ value and \citet{Calzetti2000}'s attenuation curve, we correct the emission line ratios for dust attenuation.

We derive the following ion abundance ratios: O$^{+}$/H$^{+}$ from [\oii]/H$\beta$, O$^{2+}$/H$^{+}$ from [\oiii]$\lambda$5007/H$\beta$, C$^{2+}$/O$^{2+}$ from \ciii]/\oiii], Ne$^{2+}$/O$^{2+}$ from [\neiii]/[\oiii]$\lambda$5007, Ar$^{2+}$/O$^{2+}$ from [\ariii]/[\oiii]$\lambda$5007, Si$^{2+}$/O$^{2+}$ from \siiii]$\lambda1892$/\oiii], and Fe$^{2+}$/O$^{+}$ from [\feiii]/[\oii].
These ion pairs are selected to minimise differences in their ionisation fractions, thereby reducing the need for significant ionisation corrections.
Following the models that assume different physical properties for each ionisation zone \citep[e.g.,][]{Izotov2006,Berg2021}, we adopt $T_{\mathrm{e}}$[\oiii] for O$^{2+}$, C$^{2+}$, and Ne$^{2+}$, $T_{\mathrm{e}}$[\siii] for Ar$^{2+}$ and Si$^{2+}$, and $T_{\mathrm{e}}$[\oii] for Fe$^{2+}$.
We estimate $T_{\mathrm{e}}$[\oii] and $T_{\mathrm{e}}$[\siii] from $T_{\mathrm{e}}$[\oiii] using the relations based on photoionisation models \citep{Garnett1992}, which agree with the observed $T_{\mathrm{e}}$[\oii]-$T_{\mathrm{e}}$[\oiii] and $T_{\mathrm{e}}$[\siii]-$T_{\mathrm{e}}$[\oiii] relations of $z\sim0$ star-forming galaxies \citep{Mingozzi2022}, whose sSFRs \citep{Berg2022} are comparable to our sample.
Conversely, given that the $n_{\mathrm{e}}$ value likely depends on the critical densities of the used lines rather than the ionisation potential \citep{Harikane2025b}, we revert to a simple assumption and adopt $n_{\mathrm{e}}$[\sii] for all ionisation zones.
However, the doubly-ionised abundance ratios change only negligibly even when we adopt an $n_{\mathrm{e}}$ value 30 times higher than $n_{\mathrm{e}}$[\sii], motivated by the \ciii]-based $n_{\mathrm{e}}$ reported by \citet{Topping2025}.

Since we do not resolve different ionisation zones within the \hii\ region, the elemental abundance is obtained by summing the abundances of all ionic species.
Given that the abundances of O$^{3+}$ and higher-order oxygen ions are negligible across a wide range of ionisation parameter \citep[e.g.,][]{Berg2019} due to their high ionisation potentials ($>55$ eV), we regard $12+\log(\mathrm{O/H})$ as $12+\log(\mathrm{(O^{+}+O^{2+})/H^{+}})$ as done by many previous works \citep[e.g.,][]{Izotov2006}.

We estimate element abundance ratios by calculating ionisation correction factors (ICFs).
For example, we define $\mathrm{ICF(C^{2+}/O^{2+})}$ as $\mathrm{C/O}=\mathrm{C^{2+}/O^{2+}\times ICF(C^{2+}/O^{2+})}$, and the same applies to the other ICFs.
To obtain the ICFs, we construct stellar photoionisation models based on \citet{Isobe2023c}.
We use Cloudy \citep{Ferland2013} to simulate the photoionisation of a nebula illuminated by BPASS \citep{Stanway2018} stellar population spectra with a stellar age of 10 Myr and an upper star mass cut of 100 $M_{\odot}$ under the assumption of the \citet{Salpeter1955} initial mass function (IMF).
We adopt BPASS models for binary-star populations because most of O and B type stars in the MW \citep[e.g.,][]{Duchene2013} and the Magellanic Clouds \citep[e.g.,][]{Sana2014} are likely part of binary systems, which suggest a high binary fraction for the young stellar population \citep{Stanway2018}.
We assume the stellar metallicity to be the same as that of nebular component, since massive, short-lived stars are expected to have metallicities comparable to that of the surrounding nebula.
We grid our models using O/H within $6.69\leq12+\log(\mathrm{O/H})\leq9.69$ in 0.25-dex increments and ionisation parameter $U$ within $-3.5\leq\log(U)\leq-0.5$ in 0.25-dex increments.
We assume a hydrogen density of $n_{\rm H}=300$ cm$^{-3}$ inferred from $n_{\mathrm{e}}$[\oii]\ or $n_{\mathrm{e}}$[\sii] at similar $z$ \citep[e.g.,][]{Isobe2023b,Reddy2023b,LiSijia2025,Topping2025b,Harikane2025b}, and that the He/H and metal-to-O ratios are the same as for solar abundances.

We choose the model with metallicity and [\oiii]$\lambda$5007/[\oii] values closest to the observed values.
The obtained ICFs are close to unity (Tab.~\ref{tab:neb}), indicating small corrections.
It is worth mentioning that our $\mathrm{ICF(Fe^{2+}/O^{+})}$ values are comparable to those of \citet{Rodriguez2005}, and that \citet{Mendez-Delgado2024} report that \citet{Izotov2006}'s ICFs tend to overestimate the total Fe abundance compared to \citet{Rodriguez2005}'s ICFs.

We obtain element abundance ratios of C/O, Ne/O, Ar/O, Si/O, and Fe/O from the corresponding ion abundance ratios and the ICFs.
We estimate errors of these abundance ratios as well as \met\ and the ICFs by calculating them 1000 times based on the flux values randomly perturbed by their errors.
We verify that the estimated errors appropriately account for the flux errors of the stacked spectra.
However, we note that these flux errors reflect only the measurement errors of the individual spectra and do not characterise the underlying distribution of the sample.
The nebular properties are listed in Tab.~\ref{tab:neb}.

Here, we verify the insensitivity of our results to different assumptions and methodologies.
First, we assess the potential biases introduced by altering the order of continuum subtraction, which may be non-negligible, particularly when a large fraction of the sample exhibits substantial continuum emission.
We create stacked spectra of galaxies whose continuum is modelled with the same function as for our stacks (see Section \ref{subsec:emis}) and subtracted in advance.
We confirm that this analysis changes \met\ and the chemical abundance ratios by only $\lesssim0.1$ dex, which does not impact our conclusions.
Second, we test the extinction curve of the Small Magellanic Cloud \citep[SMC;][]{Gordon2003}.
We find that the assumption of the SMC extinction curve changes \met, Ne/O, Ar/O, and Fe/O ratios by only $\lesssim0.01$ dex.
The SMC extinction curve can decrease C/O and Si/O ratios by $\sim0.1$ dex, which does not change our conclusions.
Third, we test the empirical $T_{\mathrm{e}}$[\siii]-$T_{\mathrm{e}}$[\oiii] relation based on $z\sim0$ observations \citep{Hagele2006}, which provides $T_{\mathrm{e}}$[\siii$]\simeq T_{\mathrm{e}}$[\oiii] for the $T_{\mathrm{e}}$[\oiii] values of our stacks.
We find that this relation decreases Ar/O and Si/O by $\lesssim0.07$ dex and $\lesssim0.2$ dex, respectively, yet our conclusions still hold.
Fourth, we test AGN photoionisation models \citep{Isobe2025}, considering the possibility of a hidden AGN population \citep{Geris2025}.
The AGN models can impact \met, C/O, Ar/O, and Si/O ratios by only $\lesssim0.05$ dex.
Although the AGN models can decrease Ne/O ratios and increase Fe/O ratios by $\sim0.1$ dex, they do not affect our conclusions.
Finally, we compute Cloudy photoionisation models with silicate dust depletion.
We adopt a dust-to-gas mass ratio (D/G) of $10^{-3}$, which is around 10\% of the MW \citep[e.g.,][]{Draine2004}.
The adopted D/G ratio is also the maximum of the simulated values of \citet{Popping2017} within the metallicity range of $12+\log(\mathrm{O/H})<8$.
We find that the inclusion of dust with $\mathrm{D/G}=10^{-3}$ impacts the chemical abundance ratios by only $\lesssim\pm0.01$ dex, which does not influence our conclusions.

\subsection{Si/O ratios of individual galaxies at $z>4$} \label{subsec:gnz11}

\begin{table}
	\centering
	\caption{Nebular properties of GN-z11 and RXCJ2248 derived and reanalysed by this work. $^{\dagger}$: Adopted by \citet{Alvarez2025}. $^{\ddagger}$: Adopted by \citet{Topping2024a}.}
	\label{tab:nebidv}
	\begin{tabular}{cccc}
		\hline
        Property & \multicolumn{2}{c}{GN-z11} & RXCJ2248\\
        \hline
        Assumed $\log(n_{\mathrm{e}}/\mathrm{cm^{-3}})$ & 3$^{\dagger}$ & 5 & 5$^{\ddagger}$\\
        $E(B-V)$ & $0.00^{+0.37}_{-0.00}$ & $0.00^{+0.37}_{-0.00}$ & $0.00^{+0.02}_{-0.00}$\\
        $T_{\mathrm{e}}$[\oiii]\,($10^{4}$\,K) & $1.41^{+0.19}_{-0.20}$ & $1.24^{+0.15}_{-0.16}$ & $2.65^{+0.04}_{-0.04}$\\
        \met & $7.87^{+0.19}_{-0.16}$ & $8.37^{+0.19}_{-0.15}$ & $7.41^{+0.02}_{-0.02}$\\
        ICF($\mathrm{C^{2+}/O^{2+}}$) & $1.07^{+0.01}_{-0.03}$ & $1.07^{+0.01}_{-0.03}$ & $1.97^{+0.34}_{-0.19}$\\
        $\log(\mathrm{C/O})$ & $-0.61^{+0.07}_{-0.09}$ & $-0.64^{+0.07}_{-0.08}$ & $-0.69^{+0.07}_{-0.05}$\\
        ICF($\mathrm{Ne^{2+}/O^{2+}}$) & $0.95^{+0.01}_{-0.01}$ & $0.96^{+0.01}_{-0.00}$ & $0.99^{+0.00}_{-0.00}$\\
        $\log(\mathrm{Ne/O})$ & $-0.70^{+0.07}_{-0.06}$ & $-0.72^{+0.07}_{-0.06}$ & $-0.65^{+0.01}_{-0.01}$\\
        ICF($\mathrm{Si^{2+}/O^{2+}}$) & $1.60^{+0.11}_{-0.15}$ & $2.19^{+0.20}_{-0.48}$ & $7.23^{+3.33}_{-1.77}$\\
        $\log(\mathrm{Si/O})$ & $-1.45^{+0.09}_{-0.12}$ & $-1.37^{+0.15}_{-0.06}$ & $-1.21^{+0.17}_{-0.13}$\\
        \hline
	\end{tabular}
\end{table}

As mentioned in Section \ref{sec:intro}, no Si/O measurements have been reported for emission-line galaxies at $z>4$ so far.
Currently, only RXCJ2248, a strongly-lensed galaxy at $z=6.11$ \citep[e.g.,][]{Mainali2017,Schmidt2017}, has published values of \siiii]$\lambda\lambda$1883,1892 fluxes observed with the NIRSpec R1000 grating \citep{Topping2024a}.
Additionally, we identify \siiii]$\lambda\lambda$1883,1892 detections in GN-z11, a UV-bright galaxy at $z=10.6$ \citep[e.g.,][]{Oesch2016,Bunker2023,Cameron2023}, which has a very long exposure time of up to $\sim 30$ hours in total across two sets of MSA observations with the F170LP-G235M band (\citealt{Maiolino2024a}) and two sets of integral field spectroscopy observations with F170LP-G235M \citep{Maiolino2024GN-z11_PopIII}. More details on the combination of these datasets will be provided in a separate paper (Maiolino et al. in preparation).
We use these observational results to derive Si/O ratios of GN-z11 and RXCJ2248, reanalysing other abundance ratios in the same way as we do for our stacks.

We measure fluxes of \oiii], \siiii], and \ciii] from the deep F170LP-G235M spectrum of GN-z11, following the procedure presented in Section \ref{subsec:emis}.
We take [\oiii]$\lambda$5007 and H$\alpha$ fluxes observed with the Medium Resolution Spectrograph of \textit{JWST}/MIRI reported by \citet{Alvarez2025}.
The other emission line fluxes based on the NIRSpec R1000 observations are drawn from \citet{Bunker2023} and \citet{Maiolino2024a} for GN-z11, and \citet{Topping2024a} for RXCJ2248.
Since no measurement values of [\sii]$\lambda\lambda$6716,6731 fluxes are available for GN-z11 or RXCJ2248, we first assume $n_{\mathrm{e}}=10^{3}$ cm$^{-3}$ for GN-z11 and $n_{\mathrm{e}}=10^{5}$ cm$^{-3}$ for RXCJ2248, which are the same values as those adopted in \citet{Alvarez2025} and \citet{Topping2024a}, respectively, for the sake of consistency.
In addition, we assume $n_{\mathrm{e}}=10^{5}$ cm$^{-3}$ for GN-z11, given the high $n_{\mathrm{e}}$ values derived from UV line ratios \citep{Senchyna2024,Maiolino2024a}.
Using the fixed $n_{\mathrm{e}}$ values, we derive $E(B-V)$ and $T_{\mathrm{e}}$[\oiii] values iteratively.
We obtain \met, C/O, Ne/O, and Si/O values in the same manner as Section \ref{subsec:neb}.
Note that neither GN-z11 nor RXCJ2248 has measurements of [\ariii] and [\feiii] fluxes, which prevents us from constraining Ar/O and Fe/O ratios self-consistently.
Instead, we cite the Fe/O value of GN-z11 reported by \citet{Nakane2024}, who obtain the stellar Fe/H value based on the restframe UV continuum divided by the nebular O/H value (see also e.g., \citealt{Steidel2016}).
The Fe/O value is comparable to that based on the Fe\,\textsc{ii} emission line complex \citep{Ji2025d,Nakane2025}.

The derived nebular properties are listed in Table \ref{tab:nebidv}.
We confirm that the \met\ value of GN-z11 with $n_{\mathrm{e}}=10^{5}$ cm$^{-3}$ is higher than that with $n_{\mathrm{e}}=10^{3}$ cm$^{-3}$ because the high-$n_{\mathrm{e}}$ assumption is approaching to the critical density of [\oiii]$\lambda$5007 ($6\times10^{5}$ cm$^{-3}$), which decreases the $T_{\mathrm{e}}$[\oiii] value \citep[see also][]{Hayes2025}.
However, we find that the abundance ratios based on $n_{\mathrm{e}}=10^{3}$ cm$^{-3}$ and $10^{5}$ cm$^{-3}$ are consistent with each other within the 1$\sigma$ errors.
In the following, we consider the results based on $n_{\mathrm{e}}=10^{3}$ cm$^{-3}$ for consistency with \citet{Alvarez2025}.
We confirm that both GN-z11 and RXCJ2248 have $E(B-V)=0$, indicating negligible dust attenuation, consistent with previous findings from the literature \citep[e.g.,][]{Bunker2023,Tacchella2023,Topping2024a,Yanagisawa2024,Alvarez2025}.
Our $T_{\mathrm{e}}$ values also agree with those of \citet{Alvarez2025} and \citet{Topping2024a} within $1\sigma$.
It is noteworthy that our $T_{\mathrm{e}}$, \met, and C/O values of GN-z11 are consistent with the fiducial values of \citet{Cameron2023} and \citet{Ji2025d}, who use [\neiii] to infer [\oiii]$\lambda$5007.
Our Ne/O ratio of GN-z11 is comparable to the solar Ne/O ratio of $\log(\mathrm{Ne/O})=-0.76$ \citep{Asplund2009}, which \citet{Cameron2023} adopt to derive the fiducial values.
It is worth mentioning that our ICF($\mathrm{C^{2+}/O^{2+}}$) value of RXCJ2248 is in good agreement with that of \citet{Topping2024a} based on \citet{Berg2019}'s model, validating our method of obtaining the ICFs.
We note that, although the difference remains consistent within the 1$\sigma$ errors, the measurement value of our C/O ratio is 0.14 dex higher than that of \citet{Topping2024a}.
The difference mainly originates from the difference in $\mathrm{C^{2+}/O^{2+}}$ values, suggesting different procedures to derive ion abundances.
We confirm that our method accurately reproduces the $\mathrm{C^{2+}/O^{2+}}$ values of \citet{Berg2019}.

Although we confirm our methodology as above, the large ICF($\mathrm{Si^{2+}/O^{2+}}$) value of RXCJ2248 indicates the need for a large correction to the derived Si/O compared to our stacks and GN-z11.
This is because RXCJ2248 has an extremely high [\oiii]$\lambda$5007/[\oii] ratio of 184 \citep{Topping2024a}, which suggests a dominant population of highly-ionised Si ions.
It should be noted that both GN-z11 and RXCJ2248 are NOEGs \citep{Cameron2023,Topping2024a}, suggesting anomalous chemical enrichment.
The possibility of massive or AGB stars discussed for N/O enhancement (see Section \ref{sec:intro}) may leave room for enhancement of heavy $\alpha$-elements via PISNe or Type-Ia SNe, respectively.

\subsection{Si/O ratios of local galaxies} \label{subsec:loc}
We reanalyse Si/O ratios of $z\sim0$ galaxies, which have been reported by only a few references \citep[e.g.,][]{Garnett1995b,Izotov1999}.
We collect the necessary emission line fluxes of 38 galaxies (\citealt{Berg2016,Berg2019,Berg2021}; and the references therein), which include eight galaxies with $>3\sigma$ detections of both \oiii] and \siiii].
Note that we use the fluxes already corrected for dust attenuation.
When deriving nebular properties, we follow the same procedure as we do for our stacks for consistency, except for the galaxies without [\oii]$\lambda\lambda$3726,3729.
For such galaxies, we use [\oii]$\lambda\lambda$7320,7330 to derive O$^{+}$ abundances.
It is worth mentioning that we can reproduce \citet{Garnett1995b}'s result if we adopt $T_{\mathrm{e}}$(\oiii) for Si$^{2+}$ as done by \citet{Garnett1995b}.
However, adopting $T_{\mathrm{e}}$(\siii) as we do in Section \ref{subsec:neb} increases the Si/O ratio by $\sim0.2$ dex, which highlights the importance of adopting the corresponding $T_{\mathrm{e}}$ value for each ionisation zone \citep{Berg2021}.
We derive a median value of $\log(\mathrm{Si/O})=-1.56$, which lies between those reported by \citet{Garnett1995b} and \citet{Izotov1999}, while a direct comparison with these references is not straightforward due to differences in methodology and sample.

Additionally, we reanalyse the C/O, Ne/O, Ar/O, and Fe/O ratios of the $z\sim0$ galaxies in the same way as described above.
We also take Ne/O and Ar/O ratios of $z\sim0$ metal-poor emission-line galaxies from \citet{Izotov2006} and Fe/O ratios of $z\sim0$ star-forming regions (both Galactic and extragalactic) from \citet{Mendez-Delgado2024}, all of which are based on the direct-$T_{\mathrm{e}}$ method.
Note that the majority of the $z\sim0$ galaxies with Si/O and C/O measurements are biased towards starburst galaxies with $\log(\mathrm{sSFR/Gyr^{-1}})\sim0.9$ \citep{Berg2016,Berg2019}, which is higher than that of the star-formation main sequence (SFMS) at $z\sim0$ \citep[e.g.,][]{Chang2015}.
Cross-matching the \citet{Izotov2006} catalog with the galaxy property catalog of the Max Planck Institute for Astrophysics-Johns Hopkins University (MPA-JHU) group, which provides $M_{*}$ \citep{Kauffmann2003} and SFR values \citep{Brinchmann2004}, we confirm that the \citet{Izotov2006} galaxies have a median value of $\log(\mathrm{sSFR/Gyr^{-1}})=0.56$, exceeding that of the $z\sim0$ SFMS.

\section{Results and discussions} \label{sec:res}
\begin{figure*}
    \includegraphics[width=\textwidth]{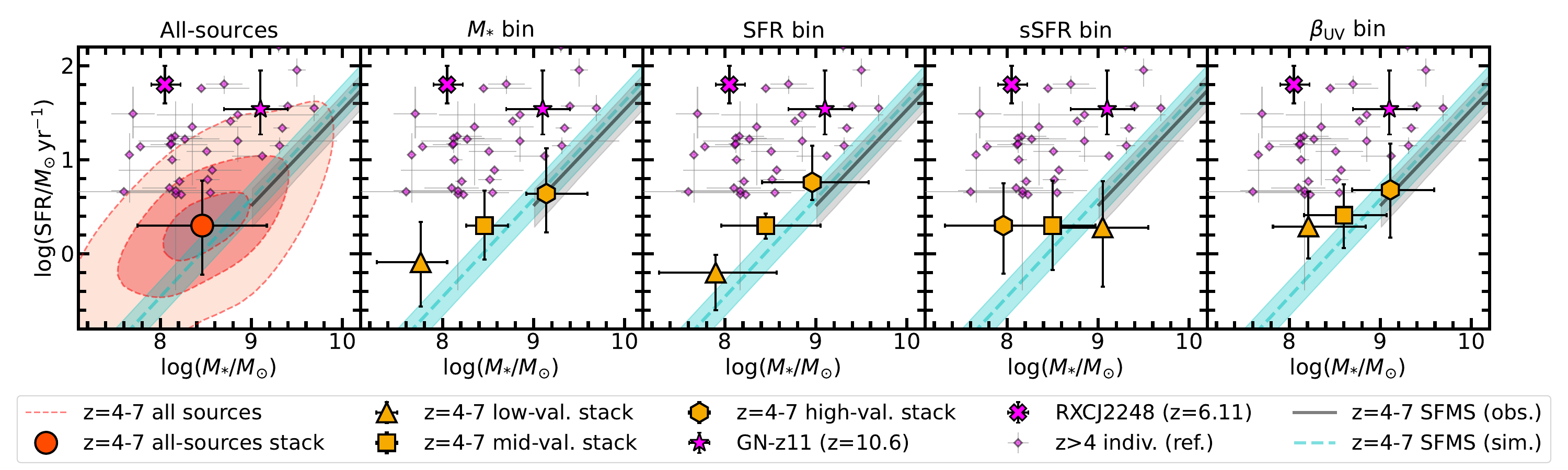}
    \caption{SFR as a function of $M_{*}$. 
    The distribution of all sources at $z=4$--7 in our sample is represented by the red contour, whose levels are 50 and 84 percentiles of the kernel density estimated with the \texttt{scipy} package \texttt{gaussian\_kde}.
    The left panel shows the all-sources stack (red circle), and the others show the low- (orange square), mid- (orange pentagon), and high-value (orange hexagon) stacks with respect to the property indicated at the top of the panel. We plot the measurements of GN-z11 (magenta star) and RXCJ2248 (magenta cross; Section \ref{subsec:gnz11}). The small magenta diamonds are the references of $z>4$ individual galaxies (see Appendix \ref{apsec:ref} for details). The grey solid line and the cyan dashed line with the shades denote the SFMS based on the $M_{*}$-complete observations \citep{Simmonds2025} and the zoom-in radiation hydrodynamics simulations of \textsc{thesan-zoom} \citep{McClymont2025a}, respectively (Section \ref{subsec:fund} for more details). The median $M_{*}$ and SFR values of our stack are generally closer to the SFMS than most of the individual galaxies.}
    \label{fig:sfms}
\end{figure*}

\begin{figure*}
    \includegraphics[width=0.99\textwidth]{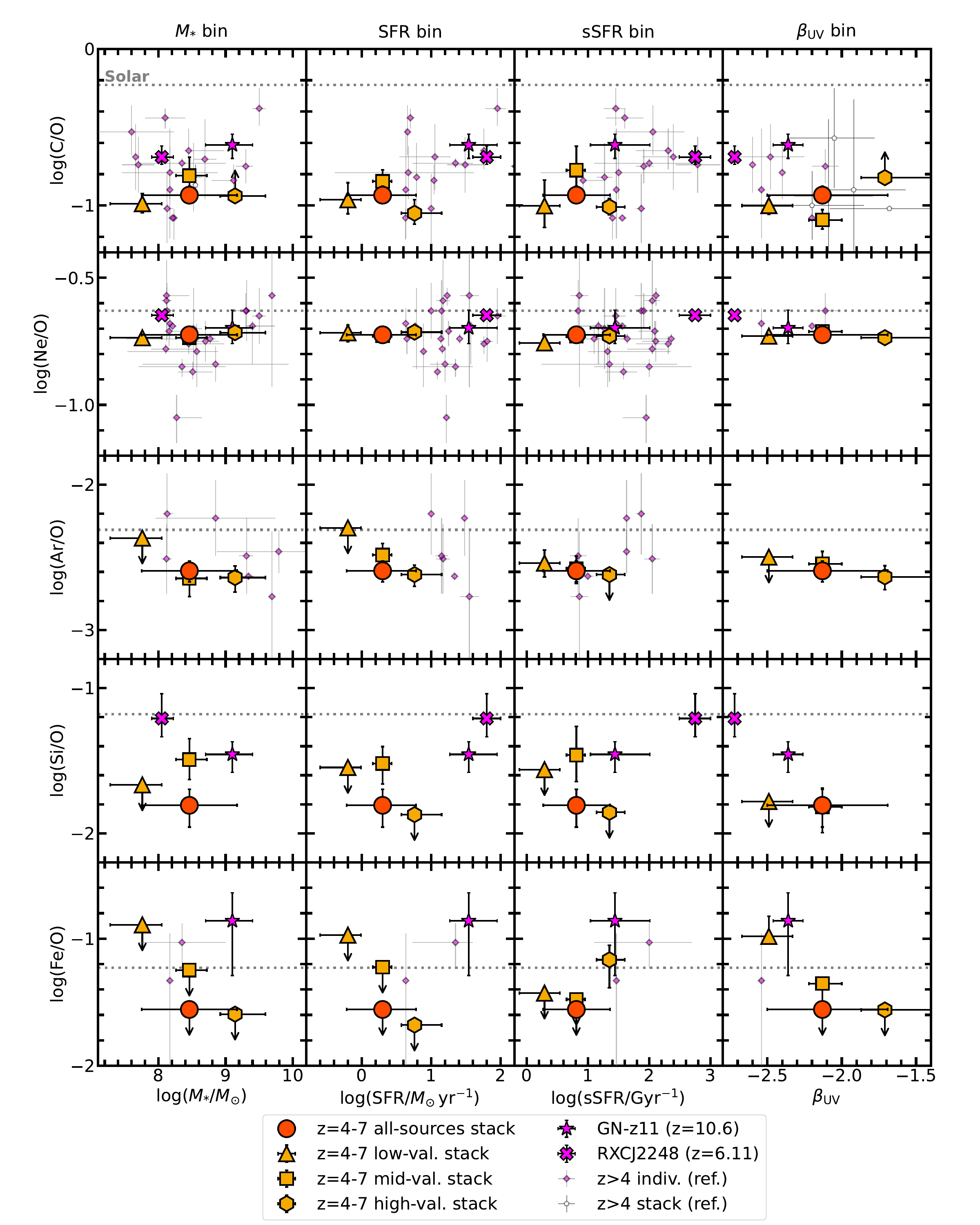}
    \caption{
    Chemical abundance ratios (C/O: top, Ne/O: second top, Ar/O: third top, Si/O: second bottom, Fe/O: bottom) as a function of $M_{*}$ (left), SFR (second left), sSFR (second right), and \buv\ (right).
    The small white pentagons are the references of $z>4$ stacks \citep{Hu2024,Hayes2025}.
    The horizontal dotted line shows the solar abundance \citep{Asplund2021}.
    The other symbols are the same as in Fig.~\ref{fig:sfms}.
    Our stacks generally show low C/O, moderate Ne/O, marginally low Ar/O, low Si/O, and low Fe/O ratios compared to individual galaxies at $z>4$, with the exception that the high-sSFR stack and the blue-\buv\ stack exhibit enhanced Fe/O ratios comparable to those of the $z>4$ galaxies.
    }
    \label{fig:abunbin}
\end{figure*}

\begin{figure*}
    \includegraphics[width=\textwidth]{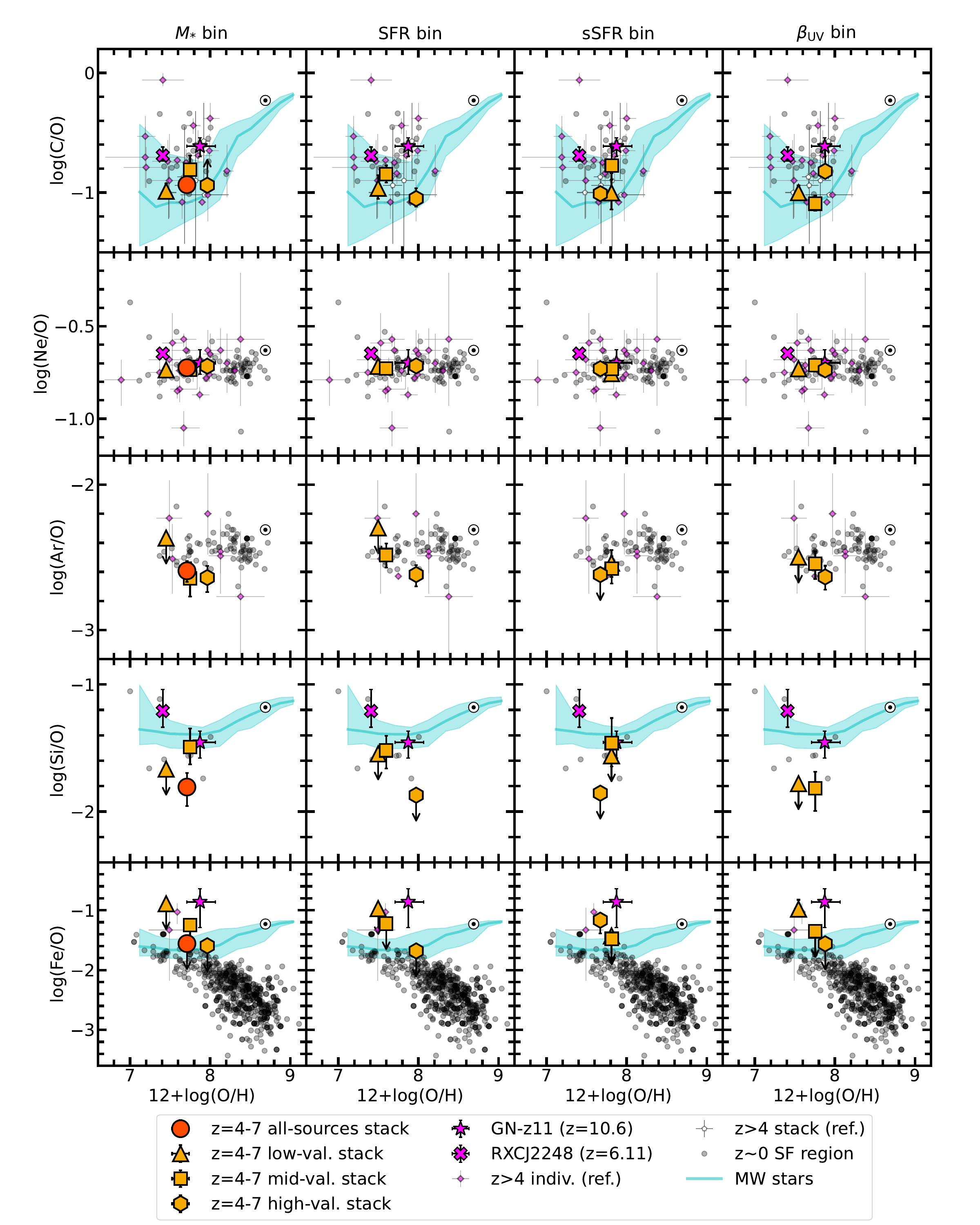}
    \caption{Same as Fig.~\ref{fig:abunbin} but chemical abundance ratios as a function of \met. We add the measurements of $z\sim0$ star-forming regions (gray dots; \citealt{Izotov2006,Mendez-Delgado2024}; Section \ref{subsec:loc}). The cyan solid curve with the shade shows the median with the 16th-84th percentile range of the distribution of the MW stars \citep[][see also Section \ref{subsec:cnear}]{Abdurrouf2022}. The circled dot denotes the solar abundance \citep{Asplund2021}. Compared to the $z\sim0$ star-forming regions and the MW stars, our stacks generally exhibit lower Si/O ratios, while the high-sSFR stack and the blue-\buv\ stack have higher Fe/O ratios.}
    \label{fig:abunmet}
\end{figure*}

\begin{figure*}
    \includegraphics[width=\textwidth]{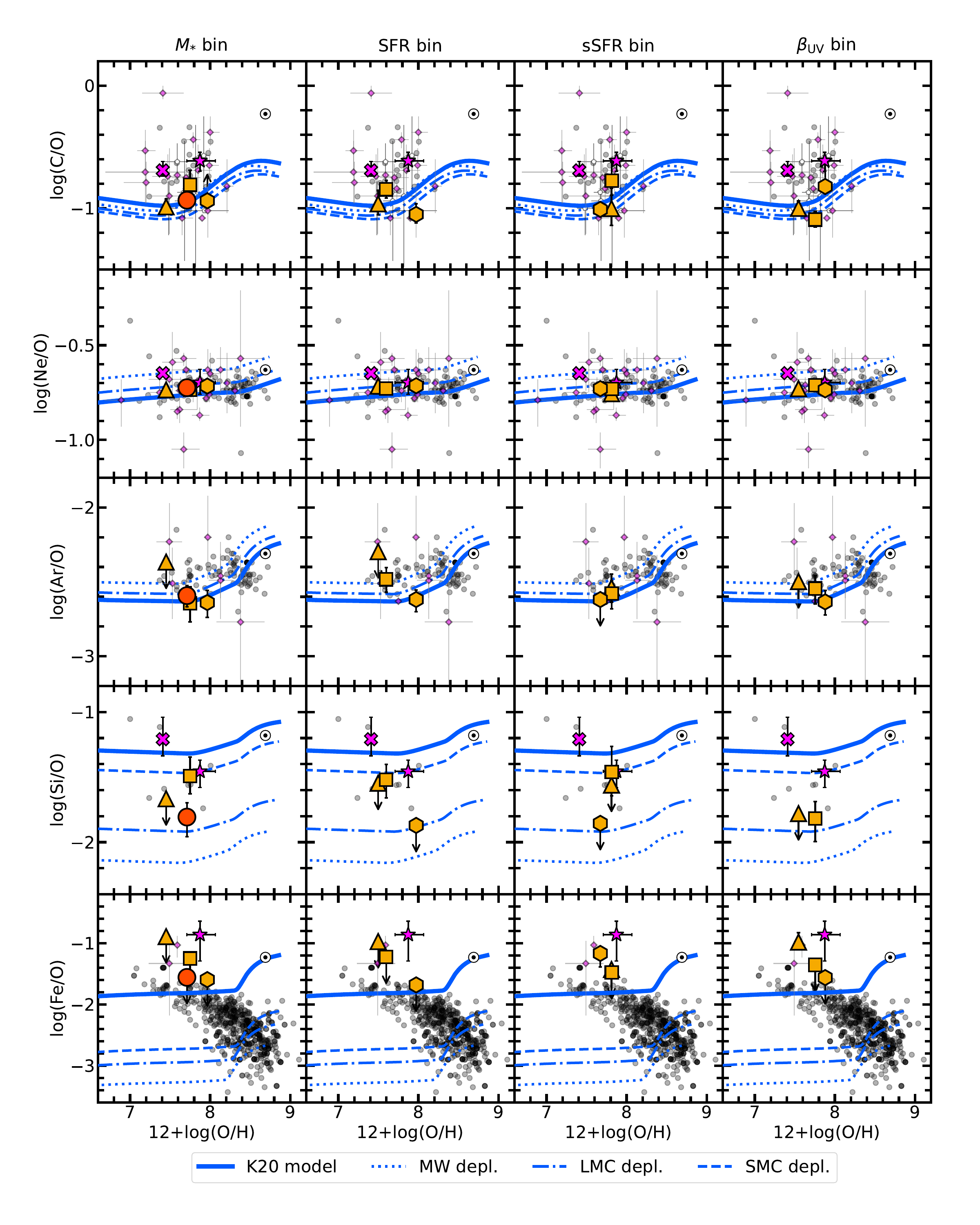}
    \caption{Same as Fig.~\ref{fig:abunmet} but we replace the MW star distribution with the MW chemical evolution model without dust depletion \citep[K20 model; blue solid curve;][see Section \ref{subsec:cnear}]{Kobayashi2020}. The K20 model changes with the assumption of dust depletion in the MW (blue dashed), the Large Magellanic Cloud (LMC; blue dashdot), and the Small Magellanic Cloud (SMC; blue dotted) with the H column density of $N_{\mathrm{H}}=10^{21}$ cm$^{-2}$ \citep{Roman-Duval2022}. The low Si/O ratio of the all-sources stack is comparable to that of the K20 model with the LMC depletion pattern.}
    \label{fig:abunmet_mod}
\end{figure*}

\subsection{Fundamental properties} \label{subsec:fund}
First, we report fundamental properties of our sample.
Fig.~\ref{fig:sfms} shows the median SFR and $M_{*}$ values of our sample of galaxies with the error bars representing the 16th-84th percentile range of the distributions.
We plot GN-z11 and RXCJ2248 from this work (Section \ref{subsec:gnz11}) and other $z>4$ individual galaxies from the literature with abundance ratios based on the direct-$T_{\mathrm{e}}$ method for accuracy (see Appendix \ref{apsec:ref} for more details).
Although measurement methods reported in the literature are heterogeneous, we give priority to citing SFR$_{10}$ or SFR estimates that trace similar timescales (i.e., from H$\alpha$ or H$\beta$).
If such SFRs are not available, we use SFRs tracing longer timescales with the implication that they serve as lower limits of SFR$_{10}$ at high $z$ on average (e.g., \citealt{Simmonds2025}; see Appendix \ref{apsec:ref} for more details).
Fig.~\ref{fig:sfms} illustrates that the median SFRs of our stacks are lower than those of the $z>4$ individual galaxies at a given $M_{*}$.
It is probably a natural consequence of requiring auroral line detections for the individual galaxies, which should be biased towards galaxies with brighter emission lines, and thus, with higher SFRs and sSFRs.
In contrast, we note that our JADES spectroscopic sample includes NIRCam-selected sources \citep{DEugenio2025}, which probe a fainter galaxy population.

Fig.~\ref{fig:sfms} shows the SFMS of the $M_{*}$-complete sample
derived at $9.0\leq\log(M_{*}/M_{\odot})\leq10.3$
based on the JADES NIRCam photometry \citep{Simmonds2025} and the \textsc{thesan-zoom} simulations \citep{McClymont2025a}, where the \textsc{thesan-zoom} simulations \citep{Kannan2025} are zoom-in simulations of the large-volume (95.5 cMpc) radiation hydrodynamics simulations of \textsc{thesan} \citep{Kannan2022,Smith2022,Garaldi2022}.
\citet{Simmonds2025} and \citet{McClymont2025a} adopt \citet{Tacchella2016}'s parametrisation of the SFMS.
We refer to the values of the SFMS parameters based on SFR$_{10}$ for consistency, and plot the grey solid line \citep{Simmonds2025} and the cyan dashed line \citep{McClymont2025a} with the median $z$ of all our 564 spectra (i.e., $z$ of the all-sources stack).
The shades represent the uncertainty of the SFMSs estimated by Monte Carlo simulations based on the uncertainties of the SFMS parameters and the $z$ distribution of the all-sources stack.
We find that the median SFR of the all-sources stack is much closer to the SFMS than most of the individual galaxies.
In this sense, the all-sources stack is expected to represent more general abundance ratios at this $z$.

\subsection{C/O, Ne/O, and Ar/O ratios} \label{subsec:cnear}

Fig.~\ref{fig:abunbin} shows C/O ratios as a function of $M_{*}$, SFR, sSFR, and \buv.
Except for the high-$M_{*}$ and red-\buv\ stacks with only the lower limits on C/O, We find that our stacks
have low C/O ratios comparable to those of other $z>4$ stacks, which is slightly lower than the majority of the $z>4$ individual galaxies.
In the $z>4$ individual galaxies, we identify a tentative positive correlation between C/O and SFR (Kendall $\tau$ of 0.33 with a $p$-value of 0.06).
In contrast,
our stacks have Ne/O ratios comparable to those of the $z>4$ individual galaxies.
Although the number of the $z>4$ individual galaxies for which Ar abundance is measured is very limited, making definitive conclusions difficult,
our Ar/O values appear to be comparable to or marginally lower than the distribution of individual galaxies.
The top three rows of Fig.~\ref{fig:abunmet} confirm these trends.

Fig.~\ref{fig:abunmet} shows the chemical abundance ratios versus metallicity, showing compilations of individual measurements of $z\sim0$ galaxies and star-forming regions based on the direct-$T_{\mathrm{e}}$ method (see Section \ref{subsec:si}).
Our Ne/O ratios are comparable to those of the $z\sim0$ galaxies, while we highlight that our C/O and Ar/O ratios are generally lower.

These individual, well studied galaxies show a scatter at $\log(\mathrm{C/O})\sim(-1)$--($-0.4$).
Chemical enrichment models of \citet{Berg2019} explain such a scatter toward high C/O by assuming multiple starbursts with C/O enhancement by AGB stars \citep[also][]{Yin2011} and SN-driven chemically differential winds that selectively blow away oxygen \citep[see also][]{Vincenzo2016,Rizzuti2024}.
Compared to the $z\sim0$ galaxies biased towards starburst galaxies (Section \ref{subsec:si}) and the individual galaxies at $z>4$ with higher SFRs, the general population of $z=4$--7 galaxies might have less chances to experience past starbursts and/or strong outflows.
This might explain why our stacks show somewhat lower C/O than individual measurements both at $z>4$ and $z\sim0$.
This interpretation could also explain slightly low Ar/O ratios of our stacks compared to the individual measurements at $z>4$ and $z\sim0$.

Fig.~\ref{fig:abunmet} compares our measurements with the distribution of the MW stars taken from \citet{Ji2025b}, which is based on the data release 17 of the Apache Point Observatory Galaxy Evolution Experiment \citep[APOGEE;][]{Abdurrouf2022}.
\citet{Ji2025b} specifically select red giant branch stars to compare to using the following criteria: surface gravity $1.5<\log(g)<3$ and effective temperature $T_{\mathrm{eff}}<5300$ K.
As stars in the thin disc are reported to exhibit a specific chemical enrichment pattern compared to the thick disc \citep[e.g.,][]{Ji2025b}, we omit stars with azimuthal velocities higher than 150 km s$^{-1}$, which is adopted by \citet{Ji2025b} to trace the thin disc.
Because globular clusters exhibit extreme light element abundance variations (for a review on the subject see: \citealt{Milone2022}), we explicitly remove these stars from our catalogue.
This is done by cross-matching GC membership catalogues from \citet{Vasiliev2021} with our APOGEE selection, stars with membership probability greater than 50\% are removed \citep{Ji2025b}.
We calculate 16th, median, and 84th percentiles of abundance ratios of the stars within each 0.2 bin of \met, and connect them along \met\ as in Fig.~\ref{fig:abunmet}.
We find that all our stacks lie on the distribution of the MW stars in the C/O-O/H diagrams.

Fig.~\ref{fig:abunmet_mod} shows \citet{Kobayashi2020}'s chemical evolution model (K20 model, hereafter) based on the code of \citet{Kobayashi2000}.
In the K20 model, low-mass stars ($\sim0.9$--8 $M_{\odot}$) evolve into AGB stars and leave WDs, some of which explode as Type-Ia SNe according to the lifetime distribution function of \citet{Kobayashi2009}.
On the other hand, massive stars ($\sim8$--50 $M_{\odot}$) end their lives as CCSNe.
The K20 model adopts the IMF of \citet{Kroupa2008} and the SFH whose SFR values are positive for 13 Gyr with a peak at $\sim3$ Gyr, which is determined to match the observed metallicity
distribution function and agrees with the observations of WDs in the solar neighborhood \citep{Tremblay2014}.
The dotted, dot-dashed, and dashed lines show the K20 model post-processed by applying the observed dust depletion patterns typical of the ISM in the MW, LMC and SMC, respectively \citep{Roman-Duval2022}. 
Fig.~\ref{fig:abunmet_mod} illustrates that the K20 model predicts plateaus of low C/O, Ar/O, Si/O, and Fe/O ratios in the low metallicity regime, mainly due to dominant CCSNe, which increase in the high metallicity regime due to the contribution from AGB stars\footnote{Although the K20 model underproduces the C/O ratio near the solar metallicity as pointed out by \citet{Kobayashi2020}, it reproduces the observed C/O–O/H relation of MW stars at $12+\log(\mathrm{O/H})\lesssim8.1$, where we compare the model with our stacks.} and Type-Ia SNe.
The Ne/O ratio of the K20 model is relatively constant mainly due to a small contribution from low-mass stars.
We find that most of our stacks have C/O, Ne/O, and Ar/O ratios in excellent agreement with those of the low-metallicity plateaus of the K20 model at a given \met, which suggests dominant CCSN yields without a significant contribution from low-mass stars as discussed for low C/O \citep[e.g.,][]{Jones2023,Stiavelli2023} and low Ar/O \citep{Bhattacharya2025,Stanton2025}.
Note that we discuss dust depletion in Section \ref{subsec:si} rather than here because Fig.~\ref{fig:abunmet_mod} shows that dust depletion can change these ratios only slightly ($\lesssim0.2$ dex).

In summary, our findings suggest that, on average, star-forming galaxies at $z=4$--7 have a chemically young gas composition with dominant CCSN yields.
Our results further imply the rarity of high-$z$ metal-poor galaxies with excessively high C/O, low Ne/O, or high Ar/O ratios, supporting scenarios that require rare events or short time scales (see Section \ref{sec:intro}).

\subsection{Si/O ratio} \label{subsec:si}
In the second bottom rows of Figs.~\ref{fig:abunbin}, \ref{fig:abunmet}, and \ref{fig:abunmet_mod}, we find that our stacks generally have lower Si/O ratios than those of the two individual high-$z$ galaxies of GN-z11 and RXCJ2248 (Section \ref{subsec:gnz11}), the $z\sim0$ galaxies, and the MW stars.
Although it is true that dominant CCSN yields can decrease the Si/O ratio to make the plateau in the low metallicity regime of the K20 model, Si/O ratios of all our bins are even lower than the plateau, suggesting that a chemically young gas composition alone cannot explain our low Si/O ratios.
Instead, Fig.~\ref{fig:abunmet_mod} illustrates that dust depletion can significantly decrease the gas-phase Si/O ratio.
For example, the dashdot curve shows dust depletion of the LMC with the H column density of $N_{\mathrm{H}}=10^{21}$ cm$^{-2}$ \citep{Roman-Duval2022}, which is comparable to that based on our $A_{V}$ values of 0.35--0.58 (Section \ref{subsec:neb}) under the assumption of the empirical $N_{\mathrm{H}}/A_{V}$ ratio for the MW \citep[e.g.,][]{Zhu2017}.
We find that the K20 model with the depletion pattern between SMC and LMC can explain the low Si/O ratios of all our stacks.
Interestingly, all our stacks have $E(B-V)=0.1$--0.2 (Table \ref{tab:neb}), while the Si/O ratios of GN-z11 and RXCJ2248 with $E(B-V)=0$ are close to those of the K20 model without dust depletion, suggesting that dust depletion causes the low Si/O ratios of our stacks.
Our findings closely resemble those reported at intermediate $z$, in that a $z\sim2$ stacked spectrum with $E(B-V)=0.2$ has $\log(\mathrm{Si/O})=-1.8$ \citep{Steidel2016}, which is lower than that of a $z\sim2$ lensed galaxy with negligible dust attenuation ($\log(\mathrm{Si/O})=-1.6$; \citealt{Berg2018}).

The low Si/O ratios of our stacks suggest Si depletion onto dust grains in general star-forming galaxies at $z=4$--7.
Given a CCSN-dominated gas composition suggested in Section \ref{subsec:cnear}, $z=4$--7 star-forming galaxies may generally undergo a rapid creation of silicate dust by CCSNe and subsequent rapid growth in the ISM \citep{Schneider2024}.
This scenario is consistent with the flattening of the average dust attenuation curve observed at $z>4.5$ \citep{Markov2025}.
Our stacks have at most $\log(\mathrm{sSFR/Gyr^{-1}})=1.35$, which corresponds to a mass doubling time of $t_{\mathrm{dbl}}\equiv1/\mathrm{sSFR}=45$ Myr.
Given that galaxies within a similar $z$ range have rising SFHs on average \citep{Simmonds2025}, a maximum stellar age ($t_{\mathrm{max}}$) of our sample is generally longer than $t_{\mathrm{dbl}}$ of 45 Myr.
This $t_{\mathrm{max}}$ value is longer than the delay time of dust enrichment by CCSNe \citep[$\lesssim30$--40 Myr;][]{Schneider2024}, which is in line with our dust depletion scenario.
On the other hand, RXCJ2248 has $t_{\mathrm{dbl}}=1.8$ Myr \citep{Topping2024a}, which might cause negligible attenuation and depletion of dust.

It is noteworthy that the dust-to-metal mass ratios (D/M) of $z\sim0$ galaxies decline steeply from $Z\sim0.5\,Z_\odot$ to $\sim0.1\,Z_\odot$ \citep[e.g.,][]{Remy2014}.
In contrast, the low Si/O ratio of the all-sources stack with $Z\sim0.1\,Z_\odot$ is consistent with the depletion pattern of the LMC with $Z\sim0.5\,Z_\odot$ \citep[e.g.,][]{Russell1992}.
This suggests that typical high-$z$ star-forming galaxies maintain higher D/M ratios than local galaxies at the same metallicity.
Supporting this, observations of $z\sim2$--4 damped Lyman-$\alpha$ absorbers \citep{DeCia2013,DeCia2016,Wiseman2017} show nearly constant D/M ratios across $Z\sim0.1$--0.5 $Z_\odot$.
At high redshift, certain mechanisms may help maintain high D/M ratios even at low metallicities, such as high dust condensation efficiencies or accretion timescales that are insensitive to metallicity \citep[e.g.,][]{Bekki2013,McKinnon2016}, as modelled by \citet{Popping2017}.

\subsection{Fe/O ratio} \label{subsec:fe}
The bottom rows of Figs.~\ref{fig:abunbin}, \ref{fig:abunmet}, and \ref{fig:abunmet_mod} show Fe/O ratios of our stacks.
We find that the Fe/O upper limit of the all-sources stack, $\sim-0.4$ dex of the solar abundance, is lower than that of GN-z11 and other $z>4$ individual galaxies.
The Fe/O upper limit is in agreement with the $z\sim0$ galaxies, the MW stars, and the K20 model under both dust-depleted and non-depleted conditions.
This result is not in contrast with our previous finding that general star-forming galaxies at $z=4$--7 have a CCSN-dominated gas composition and dust depletion.
However, the high-sSFR and blue-\buv\ stacks exhibit supersolar Fe/O ratios, which are significantly higher than the $z\sim0$ galaxies, the MW stars, and the K20 model at a given metallicity, and comparable to those of the $z>4$ individual galaxies.
Interestingly, these individual galaxies have high sSFR and blue \buv\ values comparable to the high-sSFR and blue-\buv\ stacks.
These findings suggest that galaxies with higher sSFR and bluer \buv\ values generally show Fe/O enhancement.

\subsection{Conundrum of selective Fe/O enhancement} \label{subsec:hifeo}
\begin{figure*}
    \includegraphics[width=\textwidth]{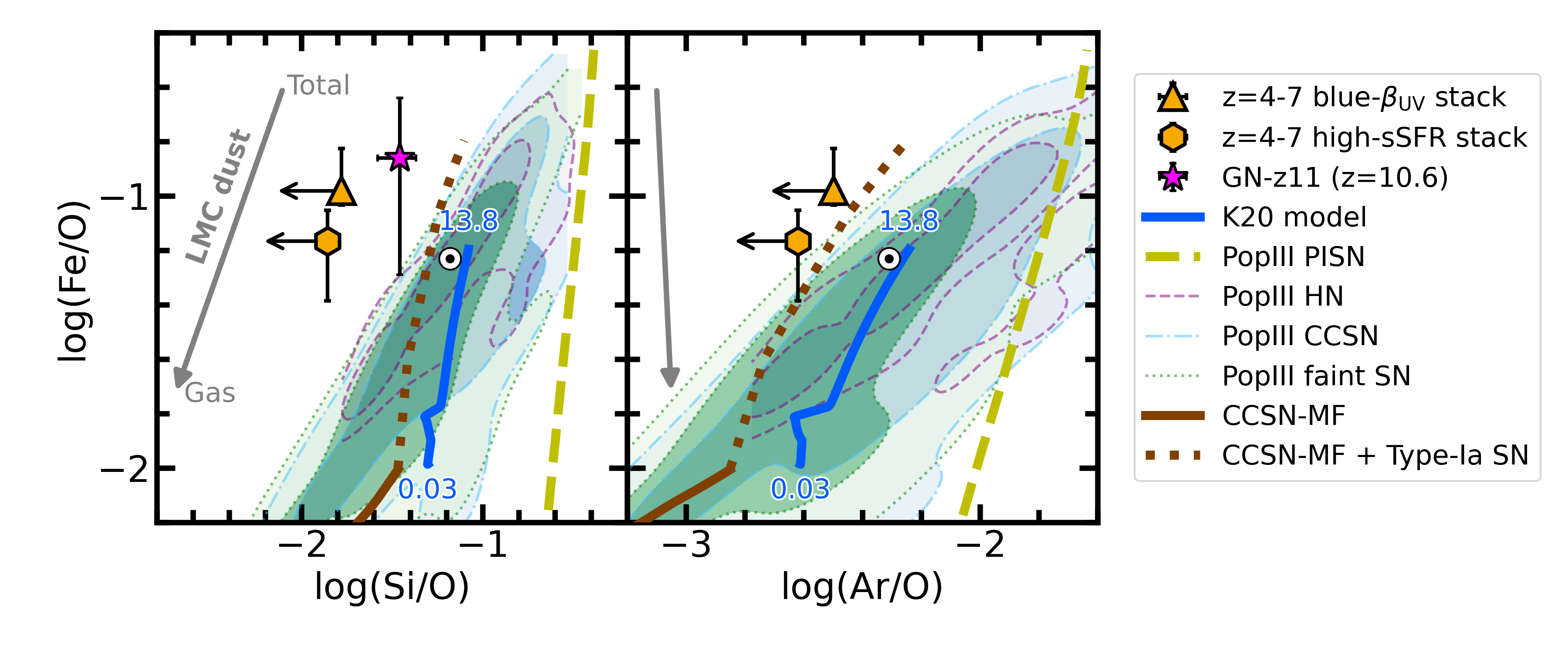}
    \caption{Fe/O ratio as a function of Si/O and Ar/O.
    The blue numbers denote the age of \citet{Kobayashi2020}'s model in Gyr.
    We add the models for Fe/O and Si/O from \citet{Vanni2024} and Ar/O from Salvadori et al. (in preparation), which are uniquely enriched by PopIII SNe with different masses  exploding as PISNe (yellow dashed curve), hypernovae (HNe; purple dashed contour), CCSNe (lightblue dashdot contour), and faint SNe (green dotted contour). The contour levels are 50 and 84 percentiles of the kernel density estimated with the \texttt{scipy} package \texttt{gaussian\_kde}.
    In the right panel, we plot \citet{Watanabe2024}'s chemical evolution model of CCSNe, which takes the mixing-and-fallback (MF) process into account (CCSN-MF; brown solid curve).
    We show \citet{Watanabe2024}'s model that mixes ejecta of CCSNe-MF and Type-Ia SNe (\citealt{Iwamoto1999}; brown dotted curve).
    The grey arrow represents the amount of dust depletion in the case of the LMC with $N_{\mathrm{H}}=10^{21}$ cm$^{-2}$ from the total abundances to the gas phase. It is difficult to satisfactorily explain the high Fe/O and low Si/O ratios of the high-sSFR stack and blue-\buv\ stack.}
    \label{fig:vanni}
\end{figure*}

\textit{What is the origin of the Fe/O enhancement?}
The Fe/O enhancement is observed in our high sSFR stack with $\log(\mathrm{sSFR/Gyr^{-1}})=1.35$ and blue \buv\ stack with $\beta_{\mathrm{UV}}=-2.49$.
The high sSFR value is equivalent to $t_{\mathrm{dbl}}=45$ Myr, which implies the onset of the starburst \citep[e.g.,][]{Mihos1994,Feulner2005,Knapen2009}.
In particular, the blue \buv\ value is accompanied by $E(B-V)=0.16$, suggesting a very blue intrinsic \buv\ of $\sim-2.7$.
This blue \buv\ can be explained by models with $\lesssim10$--20 Myr regardless of whether the nebular continuum is taken into account \citep{Cullen2024,Saxena2024}.
Such a young starburst is likely related to the Fe/O enhancement.

One possible scenario is to include exotic massive SNe such as PISNe introduced to explain Fe/O enhancements (see Section \ref{sec:intro}).
\citet{Vanni2024} have reported that supersolar Fe/O gas can be ejected not only from PISNe but also from other PopIII SNe.
The PopIII scenario obviously requires a very young, metal-poor starburst because normal Population II (PopII) stars form immediately after the metallicity exceeds $\sim10^{-4.5}$--10$^{-3}$ $Z_{\odot}$ \citep[e.g.,][]{Bromm2003,deBennassuti2017}, which is simulated to take $\sim10$ Myr after the first PopIII stars form \citep{Rusta2025}.
Note that the high-sSFR stack and blue-\buv\ stack have higher log([\oiii]$\lambda$5007/H$\beta)=0.74$ and lower log(\heii$\lambda$1640/H$\beta)=(-0.61)$--($-0.67$) (attenuation corrected) than
those simulated for galaxies dominated by PopIII stars (e.g., $>50$\% in $M_{*}$; \citealt{Rusta2025}).

Fig.~\ref{fig:vanni} shows the models of Fe/O and Si/O abundances presented by \citet{Vanni2024} and Ar/O by Salvadori et al. (in preparation), which are predicted for galaxies solely imprinted by PopIII SNe with different progenitor masses (10--100 $M_{\odot}$) and exploding with a variety of energies: PopIII hypernova (HN), CCSN, and faint SN are assumed to have explosion energies of (5--10)$\times10^{51}$, (0.9--1.5)$\times10^{51}$, and (0.3--0.6)$\times10^{51}$ erg, respectively (see \citealt{Vanni2023} for details).
These models are computed by assuming PopIII yields of \citet{Heger2010}.
We find that, although the Fe/O-Ar/O relations of the PopIII models overlap with those of the high-sSFR stack within its 1$\sigma$ error, the models predict significantly higher Si/O ratios than those of our stacks at a given Fe/O.

Another possible scenario would be Type-Ia SNe.
Although the K20 model predicts an increase in Fe/O at a higher metallicity of $12+\log(\mathrm{O/H})\gtrsim8.4$, we can increase Fe/O at a lower metallicity (or a shorter timescale) in young starbursts by assuming intermittent bursts associated with ``rejuvenation'' that recycles gas of past star formation \citep[e.g.,][]{McClymont2025c} or efficient coolants to suppress massive star formation, which results in a top-light IMF \citep[e.g.,][]{Nakane2025}.
However, Fig.~\ref{fig:vanni} illustrates that the K20 model including the ejecta of Type-Ia SNe cannot reproduce the high Fe/O, low Si/O, and low Ar/O ratios observed in the high-sSFR stack and blue-\buv\ stack simultaneously, which is because Type-Ia SNe enhance not only Fe/O but also Ar/O and Si/O ratios \citep[e.g.,][]{Iwamoto1999}.

It is worth mentioning that, to explain similar abundance patterns reported in $z\sim0$ extremely metal-poor galaxies \citep[e.g.,][]{Kojima2021}, \citet{Watanabe2024} have constructed chemical evolution models of CCSNe whose inner materials are mixed, some of which accrete to the remnant (so-called mixing-and-fallback process; e.g., \citealt{Umeda2002,Umeda2003}).
CCSNe that undergo the mixing-and-fallback process (CCSNe-MF, hereafter) can decrease heavy $\alpha$-elements, which are simply ejected by the normal CCSN model.
\citet{Watanabe2024} assume stars with 9--40 $M_{\odot}$ to follow \citet{Kroupa2001}'s IMF and to evolve into CCSNe.
The CCSN-MF yields are calculated based on the explosive nucleosynthesis code of \citet{Tominaga2007}.
\citet{Watanabe2024} adopt \citet{Ishigaki2018}'s parametrisation to express an outer boundary of mixing mass ($M_{\mathrm{mix}}$) as  $M_{\mathrm{mix}}=M_{\mathrm{cut}}+x(M_{\mathrm{CO}}-M_{\mathrm{cut}})$, where $M_{\mathrm{cut}}$ is the inner boundary of mixing mass, $M_{\mathrm{CO}}$ is the mass of the CO core, and $x$ is the mixing region factor.
\citet{Watanabe2024} mix ejecta of the CCSNe-MF and Type-Ia SNe \citep{Iwamoto1999}, which we refer to as the CCSN-MF+Type-Ia SN model hereafter.

Fig.~\ref{fig:vanni} shows the CCSN-MF+Type-Ia SN model with $x=0.2$ and metallicity $Z=0.004$, which predicts the highest Fe/O ratio of the models shown in \citet{Watanabe2024} at a given Ar/O ratio.
In addition, we plot Si/O ratios calculated with the same model.
We find that the CCSN-MF+Type-Ia SN model is closer to the observed ratios than the K20 model, which reaches the edge of the lower error bar of the observed Fe/O ratio of the high-sSFR stack.
However, the CCSN-MF+Type-Ia SN model predicts a significantly higher Si/O ratio than that of our stacks at a given Fe/O.

Here, we discuss the possibility of dust depletion and destruction.
As shown by the grey arrows in Fig.~\ref{fig:vanni}, the assumption of the LMC dust depletion would decrease the Fe/O ratios of the models more than the Si/O ratios, rendering these models even less consistent with the abundance ratios of these stacks.
It is worth noting that \citet{Curti2025b} report that a WR galaxy at $z\sim2$ has a higher gas-phase Fe/O ratio than $z\sim0$ galaxies at a given metallicity, which may point to dust displacement and sublimation in the presence of intense, localised star formation.
Although neither the high-sSFR stack nor the blue-\buv\ stack shows prominent WR features (Fig.~\ref{apfig:stkspec}), the timescale for metal accretion onto dust grains in metal-poor environments (of order several 100 Myr; e.g., \citealt{Asano2013}) may exceed the timescale over which WR features fade.
In addition, a recent computational study \citep{Hansson2025} report that
the median binding energy of Si (14.8 eV) on the surface of silicates exceeds that of Fe (6.0 eV), which may imply that Si is more readily adsorbed and relatively more resistant to sublimation.
However, it remains unclear whether conditions within \hii\ regions could achieve a temperature range in which Fe sublimates while Si remains largely unaffected.
The temperature may reach a few thousand K, as inferred from the sublimation temperature of silicates, with the exact value depending on the assumed grain size and lifetime \citep{Hansson2025}.
This question would benefit from dedicated quantitative simulations for the \hii\ region with various solid phases of iron.

It is difficult to conclude that the currently accessible models satisfactorily explain the high Fe/O and low Si/O ratios observed in the high-sSFR stack and blue-\buv\ stack.
However, the discussions above highlight the importance of further investigating supernova yields under varying progenitor characteristics and explosion scenarios, as well as considering alternative mechanisms of dust depletion and destruction.

\section{Conclusions} \label{sec:con}
We present C/O, Ne/O, Ar/O, Si/O, and Fe/O ratios of stacked spectra using 564 sources at $z=4$--7 with the \textit{JWST}/NIRSpec R1000 data from the JADES.
We perform spectral stacking with galaxy property bins of $M_{*}$, SFR, sSFR, and \buv.
The large number of high-quality spectra enables us to detect the weak emission lines of \siiii]$\lambda$1892 and [\feiii] in some of our stacks.
To the best of our knowledge, this is the first study to measure Si/O and Fe/O ratios in the general galaxy population at $z>4$.
In summary, the all-sources stack:

\begin{itemize}
    \item exhibits low C/O, moderate Ne/O, and low Ar/O ratios compared to the individual galaxies at low $z$ ($\sim0$) and high $z$ ($>4$) for a similarly low \met.
    These ratios are in very good agreement with those of the K20 chemical evolution model, which is dominated by CCSN yields in the low-metallicity regime.
    These findings suggest that general star-forming galaxies at $z=4$--7 have a chemically young gas composition with dominant CCSN yields.
    \item has a non-zero $E(B-V)=0.18$ and a Si/O ratio of $\log(\mathrm{Si/O})=-1.81$, which is lower than that of the $z\sim0$ galaxies, the MW stars, and the K20 model without dust depletion, while Si/O ratios of GN-z11 and RXCJ2248 with $E(B-V)=0$ at high $z$ are comparable to the K20 model.
    These findings suggest Si depletion on dust grains, which may be rapidly created by CCSNe.
    \item has a lower Fe/O ratio than the solar abundance by $\sim0.4$ dex.
    However, the high-sSFR stack and blue-\buv\ stack exhibit supersolar Fe/O ratios but low C/O, Ar/O, and Si/O ratios.
    The high Fe/O ratios are comparable to those of $z>4$ individual galaxies with similarly high sSFR and low \buv\ values.
    These results suggest selective Fe/O enhancement at the very early epoch of star formation.
    It is currently difficult to explain the observed abundance patterns satisfactorily, which may hint at physical processes not fully captured.
\end{itemize}

\section*{Acknowledgments}

We thank Akio Inoue, William Baker, Gareth C. Jones, Yohan Dubois, Nicholas Choustikov, Andrea Ferrara, and Koki Otaki for useful discussions.
YI and KW are supported by JSPS KAKENHI Grant No. 24KJ0202 and 24KJ1160.
RM, XJ, FDE, CS, and JS acknowledge support by the Science and Technology Facilities Council (STFC), by the ERC through Advanced Grant 695671 ``QUENCH'', and by the UKRI Frontier Research grant RISEandFALL.
RM also acknowledges funding from a research professorship from the Royal Society.
IJ acknowledges support by the Huo Family Foundation through a P.C. Ho PhD Studentship.
AS, AJB, and JC acknowledge funding from the "FirstGalaxies" Advanced Grant from the European Research Council (ERC) under the European Union’s Horizon 2020 research and innovation programme (Grant agreement No. 789056).
JW gratefully acknowledges support from the Cosmic Dawn Center through the DAWN Fellowship. The Cosmic Dawn Center (DAWN) is funded by the Danish National Research Foundation under grant No. 140.
CK acknowledges funding from the UK Science and Technology
Facility Council through grant ST/Y001443/1.
IV and SS acknowledge support by the ERC Starting Grant NEFERTITI H2020/804240.
SM and VB acknowledge support from the Leverhulme Trust Research Project Grant RPG-2021-205: ``The Faint Universe Made Visible with Machine Learning''.
AF acknowledges the support from project ``VLT- MOONS'' CRAM 1.05.03.07, INAF Large Grant 2022 ``The metal circle: a new sharp view of the baryon cycle up to Cosmic Dawn with the latest generation IFU facilities'' and INAF Large Grant 2022 ``Dual and binary SMBH in the multi-messenger era''.
WM thanks the Science and Technology Facilities Council
(STFC) Center for Doctoral Training (CDT) in Data Intensive Science at the University of Cambridge (STFC grant number 2742968) for a PhD studentship.
WM and ST acknowledge support by the Royal Society Research Grant G125142.
H\"U acknowledges funding by the European Union (ERC APEX, 101164796). Views and opinions expressed are however those of the authors only and do not necessarily reflect those of the European Union or the European Research Council Executive Agency. Neither the European Union nor the granting authority can be held responsible for them.
ECL acknowledges support of an STFC Webb Fellowship (ST/W001438/1).
BER acknowledges support from the NIRCam Science Team contract to the University of Arizona, NAS5-02015, and \textit{JWST} Program 3215.
The research of CCW is supported by NOIRLab, which is managed by the Association of Universities for Research in Astronomy (AURA) under a cooperative agreement with the National Science Foundation.
This work is based on observations made with the NASA/ESA/CSA James Webb Space Telescope. The data were obtained from the Mikulski Archive for Space Telescopes at the Space Telescope Science Institute, which is operated by the Association of Universities for Research in Astronomy, Inc., under NASA contract NAS 5-03127 for \textit{JWST}. These observations are associated with programmes \#1180, 1181, 1210, 1286, 1287, and 3215.
The authors acknowledge use of the lux supercomputer at UC Santa Cruz, funded by NSF MRI grant AST 1828315.

\section*{Data Availability}

The bulk of the \textit{JWST}/NIRSpec data used in this paper are released by the JADES NIRSpec DR1 \citep{Bunker2024} and DR3 \citep{DEugenio2025}, which are available on the JADES MAST website (\url{https://archive.stsci.edu/hlsp/jades}; MAST DOI: \href{https://dx.doi.org/10.17909/8tdj-8n28}{10.17909/8tdj-8n28}).
The rest of datasets will also be public in the MAST archive.
Our analysed data will be made available upon reasonable request.



\bibliographystyle{mnras}
\bibliography{example} 




\appendix

\section{References of the observational results} \label{apsec:ref}

In Figs.~\ref{fig:sfms}--\ref{fig:abunmet_mod}, we plot the following references of the $z>4$ individual galaxies without overlaps: \citet{
Arellano2025a,
Bhattacharya2025,
Cameron2024,
Cullen2025,
Curti2025,
Hsiao2024c,
Hsiao2024a,
Isobe2023c,
Ji2024,
Ji2025b,
Marques-Chaves2024,
Napolitano2024b,
Navarro-Carrera2024,
Schaerer2024,
Stanton2025,
Stiavelli2023,
Stiavelli2025,
Topping2025,
Zhang2025}.
These references have measurement values of \met\ and one of the following abundance ratios (C/O, Ne/O, and Ar/O) based on the direct-$T_{\mathrm{e}}$ method.
Some of these galaxies have measurement values of Fe/O based on the high-ionisation Fe line \citep{Tacchella2025} and the UV stellar continuum \citep{Nakane2025}, which we cite in this paper.
We plot \met, C/O, Ne/O, and Si/O values of GN-z11 and RXCJ2248 from this work, checking the consistency with the literature \citep[][see Section \ref{subsec:gnz11}]{Cameron2023,Alvarez2025,Topping2024a}.
The Fe/O ratio of GN-z11 is drawn from \citet{Nakane2024}, which is based on the UV stellar continuum.

We plot $M_{*}$, SFR, and \buv\ values of the $z>4$ individual galaxies taken from the following literature: \citet{
Arellano2025a,
Barchiesi2023,
Bhattacharya2025,
Bunker2023,
Cullen2025,
Curti2025,
Hsiao2024b,
Hsiao2024a,
Marques-Chaves2024,
Nakajima2023,
Napolitano2024b,
Navarro-Carrera2024,
Schaerer2024,
Stanton2025,
Stiavelli2023,
Stiavelli2025,
Tacchella2023,
Tacchella2025,
Topping2024a,
Topping2025,
Ubler2023,
Zhang2025}.
For consistency with our SFRs, we use SFR$_{10}$ values if they are available.
If SFR$_{10}$ is not available, we instead adopt the SFR measured from the Balmer lines (SFR$_\mathrm{Bal}$), as SFR$_\mathrm{Bal}$ has been reported to trace star formation on a timescale of $\sim10$ Myr \citep[e.g.,][]{McClymont2025a}, and is therefore expected to show only minor deviations from SFR$_{10}$.
The remaining available estimates are SFR$_{100}$ or SFR measured from the UV continuum (SFR$_\mathrm{UV}$), which is reported to trace a 20--30 Myr timescale of star formation \citep[e.g.,][]{McClymont2025a}.
On average, the ratio SFR$_{10}$/SFR$_{100}$ has been reported to exceed unity at high $z$ \citep[e.g.,][]{Simmonds2025}, which has been interpreted as evidence that high-$z$ galaxies exhibit more bursty star formation and rising SFHs.
Following this interpretation, a very approximate inference suggests that the actual SFR$_{10}$ values of individual galaxies would be comparable to or greater than the values that we plot in Fig.~\ref{fig:sfms}.

\section{Statistical meaning of our stacking procedure} \label{apsec:stat}
In principle, because the individual spectra exhibit different line profiles and are resampled accordingly, there is no simple formalism that directly relates the line ratios of the stacked spectrum to those of the individual spectra.
However, when the line profiles are broadly similar, particularly near the median of the sample distribution, and given that \texttt{spectres} preserves integrated flux during the resampling by design \citep{Carnall2017}, the line flux in the stacked spectrum can be approximated as the median of the individual line fluxes divided by the individual [\oiii] fluxes.

Under this approximation, the [\oiii] and H$\beta$ fluxes of the stack approach $\sim1$ and the median of the individual H$\beta$/[\oiii] ratios, respectively.
Consequently, the [\oiii]/H$\beta$ ratio of the stacked spectrum approaches the median of the individual [\oiii]/H$\beta$ ratios. 
Indeed, the median [\oiii]/H$\beta$ value of our sample is 5.79, which is only a 2\% deviation from the value of 5.70 measured in the all-sources stack.

\section{Stacked spectra in galaxy property bins} \label{apsec:stkspec}
\begin{figure*}
	\centering
    \includegraphics[width=\textwidth]{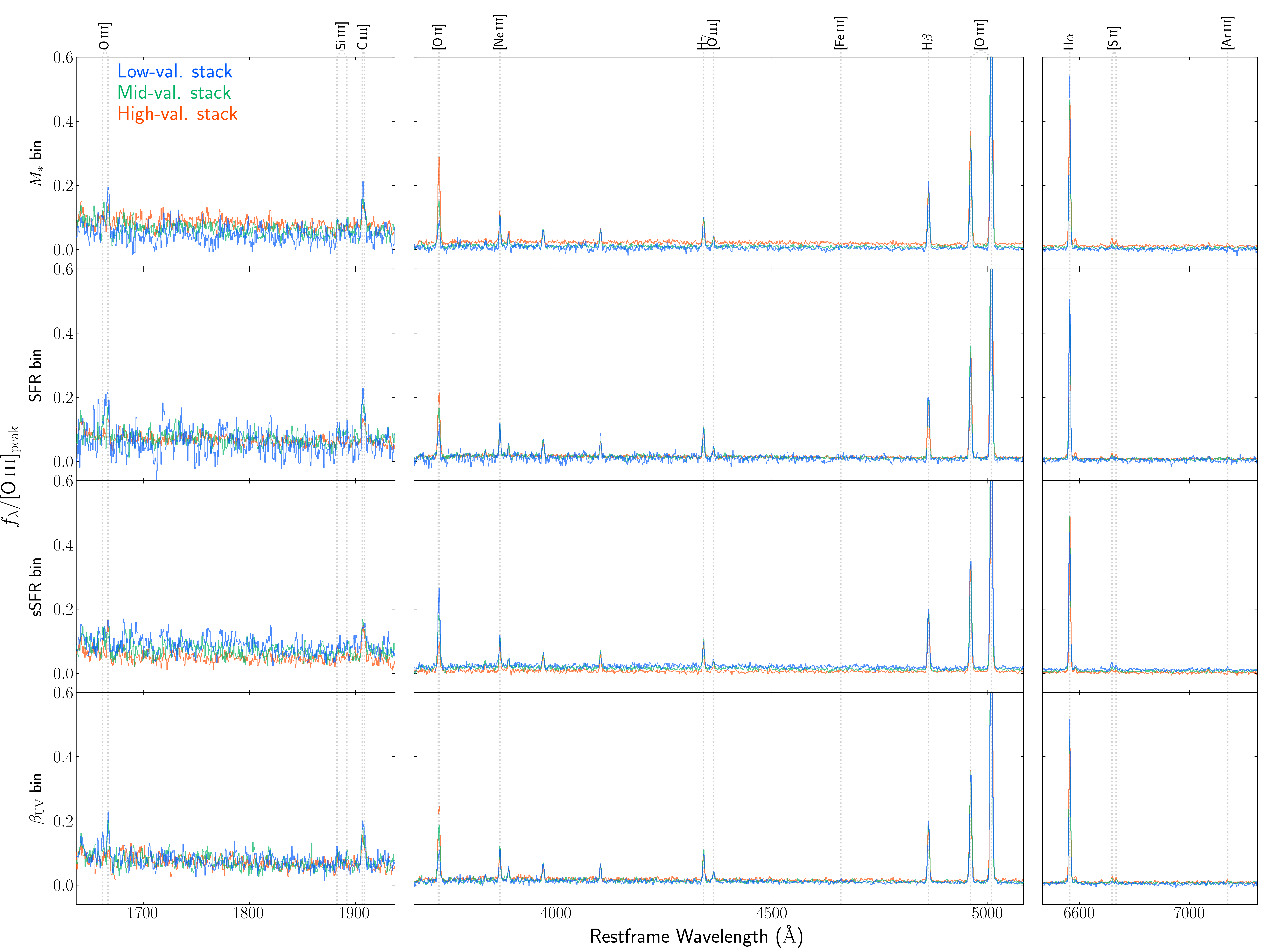}
    \caption{Same as Fig.~\ref{fig:stkspec} but for the low-value stacks (blue), mid-value stacks (green), and high-value stacks (red) in different galaxy property bins indicated at the left of each row (Section \ref{subsec:stk}).}
    \label{apfig:stkspec}
\end{figure*}

Fig.~\ref{apfig:stkspec} show our stacked spectra in different galaxy property bins (see Section \ref{subsec:stk}).
One clear trend is that the stacks with lower $M_{*}$, lower SFR, higher sSFR, and bluer \buv\ exhibit weaker [\oii]/[\oiii]$_{\mathrm{peak}}$, indicating higher [\oiii]$\lambda$5007/[\oii] values and harder ionisations.
Another trend is that the stacks with lower $M_{*}$ and higher sSFR show lower levels of continuum/[\oiii]$_{\mathrm{peak}}$, which reflect their higher sSFR values.

\section{References of the atomic data} \label{apsec:atom}

\begin{table*}
	\centering
	\caption{References of transition probabilities and collision strengths used in this paper. The abbreviations of ``Re'' and ``CE'' correspond to recombination and collisional excitation, respectively.}
	\label{tab:atom}
	\begin{tabular}{lccc}
		\hline
        Ion & Emission process & Transition probability & Collision Strength\\
        \hline
		H$^{0}$ & Re & \citet{Storey1995} & $\cdots$ \\
		C$^{2+}$ & CE & \citet{Wiese1996} & \citet{Berrington1985} \\
		O$^{+}$ & CE & \citet{FroeseFischer2004} & \citet{Kisielius2009} \\
		O$^{2+}$ & CE & \citet{FroeseFischer2004} & \citet{Lennon1994} \\
		Ne$^{2+}$ & CE & \citet{FroeseFischer2004} & \citet{McLaughlin2000} \\
		Si$^{2+}$ & CE & \citet{Ojha1988} & \citet{Dufton1989} \\
		S$^{+}$ & CE & \citet{Rynkun2019} & \citet{Tayal2010} \\
		Ar$^{2+}$ & CE & \citet{MunozBurgos2009} & \citet{MunozBurgos2009} \\
		Fe$^{2+}$ & CE & \citet{Quinet1996,Johansson2000} & \citet{Zhang1996} \\
        \hline
	\end{tabular}
\end{table*}

Table \ref{tab:atom} shows the atomic data used in this paper.
The data of H$^{0}$, C$^{2+}$, O$^{+}$, and O$^{2+}$ are the same as those adopted in \citet{Isobe2025}, and the data of Ne$^{2+}$, S$^{+}$, Ar$^{2+}$, and Fe$^{2+}$ are the same as in \citet{Isobe2022,Isobe2023c}.

\section*{Affiliations}
\textit{
$^{1}$Kavli Institute for Cosmology, University of Cambridge, Madingley Road, Cambridge, CB3 0HA, UK\\
$^{2}$Cavendish Laboratory, University of Cambridge, 19 JJ Thomson Avenue, Cambridge, CB3 0HE, UK\\
$^{3}$Waseda Research Institute for Science and Engineering, Faculty of Science and Engineering, Waseda University, 3-4-1, Okubo, Shinjuku, Tokyo 169-8555, Japan\\
$^{4}$Department of Physics and Astronomy, University College London, Gower Street, London WC1E 6BT, UK\\
$^{5}$Department of Physics, University of Oxford, Denys Wilkinson Building, Keble Road, Oxford OX1 3RH, UK\\
$^{6}$Cosmic Dawn Center (DAWN), Copenhagen, Denmark\\
$^{7}$Niels Bohr Institute, University of Copenhagen, Jagtvej 128, DK-2200, Copenhagen, Denmark\\
$^{8}$Centre for Astrophysics Research, Department of Physics, Astronomy and Mathematics, University of Hertfordshire, Hatfield, AL10 9AB, UK\\
$^{9}$Dipartimento di Fisica e Astrofisica, Universit\`a degli Studi di Firenze, Via G. Sansone 1, I-50019, Sesto Fiorentino, Italy\\
$^{10}$INAF/Osservatorio Astrofisico di Arcetri, Largo E. Fermi 5, 50125, Firenze, Italy\\
$^{11}$Department of Astronomical Science, SOKENDAI (The Graduate University for Advanced Studies), 2-21-1 Osawa, Mitaka, Tokyo, 181-8588, Japan\\
$^{12}$National Astronomical Observatory of Japan, 2-21-1 Osawa, Mitaka, Tokyo, 181-8588, Japan\\
$^{13}$Center for Interdisciplinary Exploration and Research in Astrophysics (CIERA), Northwestern University, 1800 Sherman Avenue, Evanston, IL 60201, USA\\
$^{14}$New Mexico State University, Department of Astronomy, 1320 Frenger Mall, Las Cruces, NM 88003-8001, USA\\
$^{15}$Institute of Astronomy, University of Cambridge, Madingley Road, Cambridge CB3 0HA, UK\\
$^{16}$European Southern Observatory, Karl-Schwarzschild-Strasse 2, 85748 Garching, Germany\\
$^{17}$Max-Planck-Institut f\"ur extraterrestrische Physik (MPE), Gie{\ss}enbachstra{\ss}e 1, 85748 Garching, Germany\\
$^{18}$Sorbonne Universit\'e, CNRS, UMR 7095, Institut d'Astrophysique de Paris, 98 bis bd Arago, 75014 Paris, France\\
$^{19}$AURA for European Space Agency, Space Telescope Science Institute, 3700 San Martin Drive. Baltimore, MD, 21218, USA\\
$^{20}$Space Telescope Science Institute, 3700 San Martin Drive, Baltimore, Maryland 21218, USA\\
$^{21}$Department of Astronomy and Astrophysics University of California, Santa Cruz, 1156 High Street, Santa Cruz CA 96054, USA\\
$^{22}$NSF National Optical-Infrared Astronomy Research Laboratory, 950 North Cherry Avenue, Tucson, AZ 85719, USA\\
$^{23}$NRC Herzberg, 5071 West Saanich Rd, Victoria, BC V9E 2E7, Canada
}

\bsp	
\label{lastpage}
\end{document}